\documentclass[prd,a4paper,aps,twocolumn,nofootinbib,nobibnotes,preprintnumbers,superscriptaddress]{revtex4} 
\pdfoutput=1
\usepackage[dvips]{graphicx}
\usepackage{amsmath,amssymb,mathrsfs}
\usepackage{bm}
\usepackage{times}
\usepackage{epsfig}
\usepackage{verbatim}
\usepackage{bm}
\usepackage[utf8]{inputenc}
\usepackage{graphics}
\usepackage{graphicx,epsfig,amssymb,amsmath,color, cancel}
\usepackage{soul}
\usepackage[normalem]{ulem}
\usepackage{slashed}
\usepackage{comment}

\usepackage[english]{babel}
\usepackage{amsfonts,amsmath,amssymb,amsthm,mathtools}
\usepackage{comment,enumerate,footnote,graphicx,subfloat,relsize}
\usepackage{array,tabularx,tabu,multirow,framed,afterpage}
\usepackage[usenames,dvipsnames]{xcolor}
\usepackage{cleveref}
\crefformat{pluralequation}{#2\black{eqs.~(}#1\black{)}#3}
\Crefformat{pluralequation}{#2\black{Equations~(}#1\black{)}#3}
\crefformat{pluralfigure}{#2\black{figs.~}#1#3}
\Crefformat{pluralfigure}{#2\black{Figures~}#1#3}

\newcommand{\be}{\begin{equation}}
\newcommand{\ee}{\end{equation}}

\preprint{ULB-TH/21-09}

\begin{document}
\title{
Baryogenesis via relativistic bubble expansion
}

\author{Iason Baldes}
\affiliation{Service de Physique Th\'eorique, Universit\'e Libre de Bruxelles, Boulevard du Triomphe, CP225, B-1050 Brussels, Belgium}

\author{Simone Blasi}
\affiliation{Theoretische Natuurkunde and IIHE/ELEM, Vrije Universiteit Brussel, \& The  International Solvay Institutes, Pleinlaan 2, B-1050 Brussels, Belgium}

\author{Alberto Mariotti}
\affiliation{Theoretische Natuurkunde and IIHE/ELEM, Vrije Universiteit Brussel, \& The  International Solvay Institutes, Pleinlaan 2, B-1050 Brussels, Belgium}

\author{Alexander Sevrin}
\affiliation{Theoretische Natuurkunde, Vrije Universiteit Brussel \& The International Solvay Institutes, Pleinlaan 2, B-1050 Brussels, Belgium}

\author{Kevin Turbang}
\affiliation{Theoretische Natuurkunde, Vrije Universiteit Brussel, Pleinlaan 2, B-1050 Brussels, Belgium}
\affiliation{Universiteit Antwerpen, Prinsstraat 13, B-2000 Antwerpen, Belgium}


\begin{abstract}
We present a novel baryogenesis mechanism in which the asymmetry is sourced from heavy particles which either gain their mass or are created during bubble expansion in a strong first order phase transition. These particles then decay in a CP and baryon number violating way inside the bubble. The particles are inherently out-of-equilibrium and sufficiently dilute after wall crossing so the third Sakharov condition is easily met. Washout is avoided provided the reheat temperature is sufficiently below the scale of the heavy particles. The mechanism relies on moderate supercooling and relativistic walls which --- in contrast to electroweak baryogenesis --- generically leads to a sizable gravitational wave signal, although in the simplest realisations at frequencies beyond upcoming detectors. We present a simple example model and discuss the restrictions on the parameter space for the mechanism to be successful. We find that high reheat temperatures  $T_{\rm RH} \gtrsim 10^{10}$ GeV are generally preferred, whereas stronger supercooling allows for temperatures as low as $T_{\rm RH} \sim 10^{6}$ GeV, provided the vacuum energy density is sufficiently suppressed. We briefly comment on using resonantly enhanced CP violation to achieve even lower scales.

\end{abstract}

\maketitle

\section{Introduction}
The possibility of a first order early universe phase transition~\cite{Dolan:1973qd,Weinberg:1974hy,Linde:1980tt,Linde:1981zj}, together with associated implications for inflation~\cite{Guth:1980zm,Spolyar:2011nc}, baryogenesis~\cite{Shaposhnikov:1987tw,Cohen:1990it}, dark matter~\cite{Witten:1984rs,Hambye:2018qjv}, primordial black holes~\cite{Konoplich:1999qq}, gravitational waves~\cite{Witten:1984rs,Hogan,Grojean:2006bp,Baldes:2018emh,Craig:2020jfv}, and the like, has --- over the last few decades --- captured a part of the speculative imagination of the theoretical community.

Consider a phase transition in which a scalar field, $\phi$, gains a vev, $v_{\phi}$, at some small nucleation temperature, $T_{n} \ll v_{\phi}$. Through an interaction term such as
	\begin{equation}
	\mathcal{L} \supset -\frac{\lambda}{2} \phi^{2}|\Delta|^{2},
	\end{equation}
approximately massless particles (here a scalar $\Delta$) in the symmetric phase may gain some large mass, $M_{\Delta} \sim v_{\phi} \gg T_{n}$ across the bubble wall. This is kinematically allowed provided that the bubble wall has a sufficiently high Lorentz factor, $\gamma_{\rm w}$. This can most easily be seen in the wall frame, in which the radiation in the plasma frame has energy $\sim \gamma_{\rm w} T_{n}$, so the massless $\Delta$ quanta can enter the bubble provided $\gamma_{\rm w} > M_{\Delta}/T_{n}$. The distribution of particles gaining a mass is pushed out-of-equilibrium. Their number density, $n_{\Delta}$, will not then carry the usual Boltzmann suppression, $\mathrm{Exp}[-M_{\Delta}/T_n]$, in the broken phase (until, of course, they decay or annihilate). Indeed, in this kinematic regime, the reflection probability is completely negligible so that essentially all the symmetric phase particles enter the bubble~\cite{Bodeker:2009qy,Bodeker:2017cim}. Kinematics dictates their Lorentz factor in the plasma frame to be $\gamma_{\Delta} \sim M_{\Delta}/T_{n}$ inside the bubble.

Recently, Azatov and Vanvlasselaer pointed out another novel and intriguing effect~\cite{Vanvlasselaer:2020niz}. Heavy particles receiving small corrections to their mass during a phase transition can also be produced by the light quanta being swept up by the bubble walls, provided the latter are sufficiently relativistic. Consider now the case in which $\Delta$ is already heavy in the symmetric phase. Its mass and coupling to $\phi$ can be written as,
	\begin{equation}
	\mathcal{L} \supset -\frac{1}{2}\lambda\phi^{2}|\Delta|^{2} - M_{\Delta}^{2}|\Delta|^2,
	\end{equation} 
where $M_{\Delta} \gg \lambda v_{\phi}$. Then Azatov and Vanvlasselaer showed the production $\phi \to \Delta\Delta^*$ occurs with a probability~\cite{Vanvlasselaer:2020niz}
	\begin{equation}
	P(\phi \to \Delta\Delta^*) \approx \frac{ g_{\Delta}\lambda^2 v_{\phi}^2 }{ 96 \pi^2 M_{\Delta}^{2}},
	\end{equation}
where $g_{\Delta}$ are the $\Delta$ degrees of freedom, provided $\gamma_{\rm w} T_{n} > M_{\Delta}^{2}L_{\rm w}$, which is known as the anti-adiabatic regime. Here $L_{\rm w} \approx 1/v_{\phi}$ is the wall width. Remarkably, for such relativistic walls, there is no Boltzmann type suppression of the resulting heavy particle species in the broken phase. (Between the kinematic threshold and the above, $M_{\Delta} < \gamma_{\rm w} T < M_{\Delta}^{2}L_{\rm w}$, there is an additional exponential suppression factor~\cite{Vanvlasselaer:2020niz}. We also refer the reader to~\cite{Vanvlasselaer:2020niz} for further examples including fermions.) The above production mechanism was further explored by Azatov, Vanvlasselaer, and Yin as a way of setting the DM relic abundance in~\cite{Azatov:2021ifm}. Note that also in this case, the produced $\Delta$ are out-of-equilibrium, with $\gamma_{\Delta} \sim M_{\Delta}/T_{n}$ in the plasma frame.  

\begin{figure*}[t]
\begin{center}
\includegraphics[width=200pt]{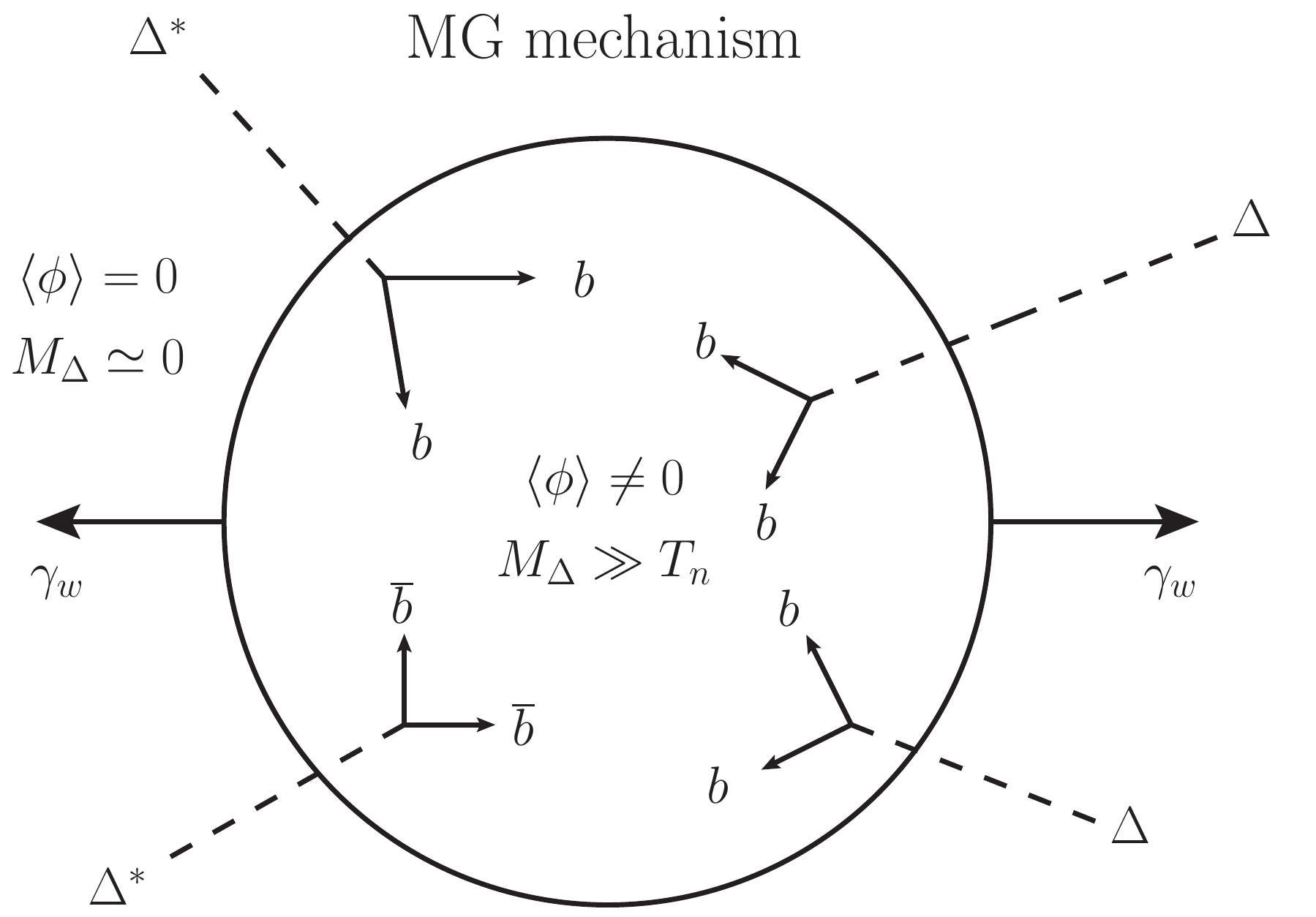} \qquad \qquad \quad \quad
\includegraphics[width=200pt]{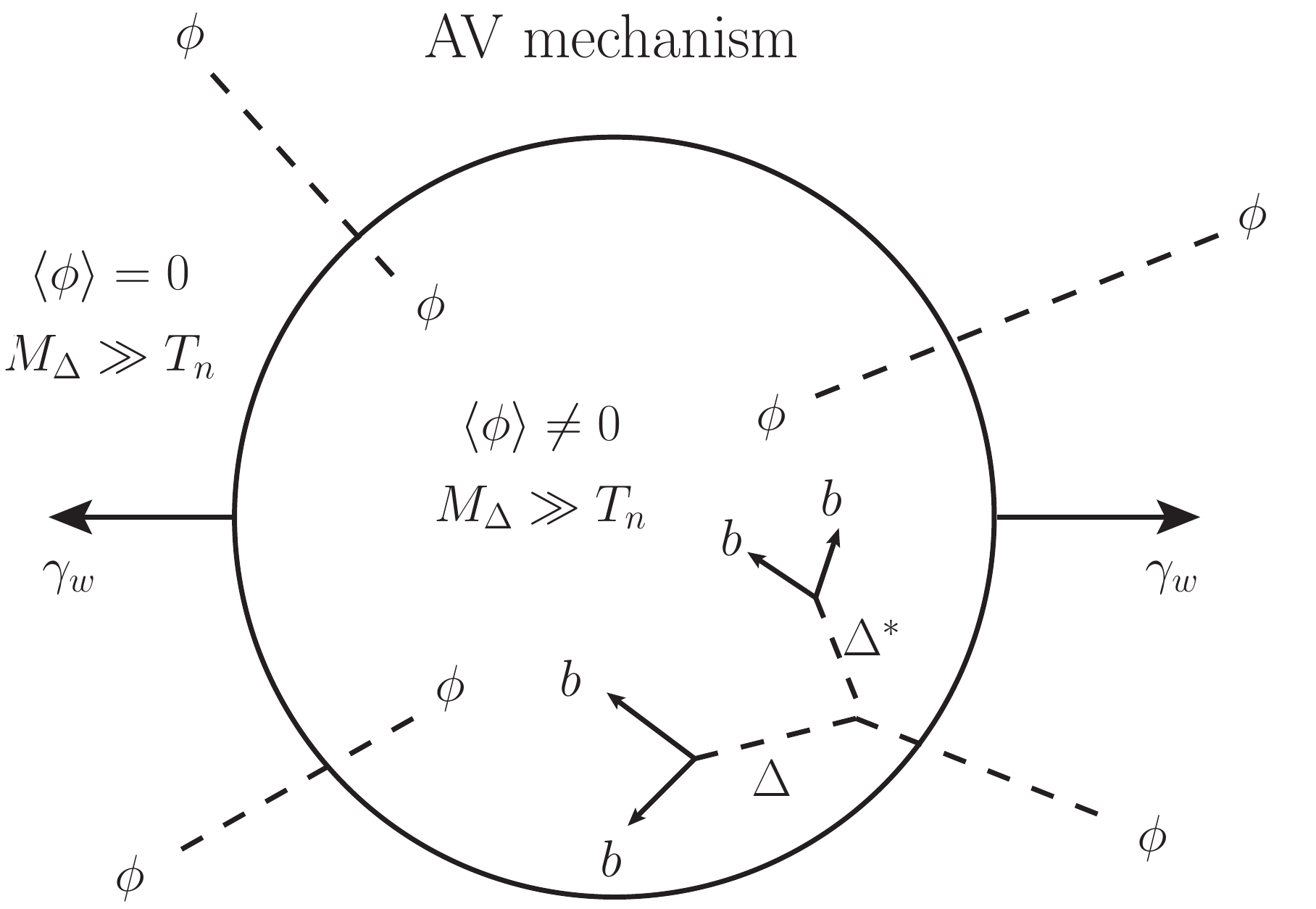}
\end{center}
\caption{Left: A sketch of the mass gain mechanism baryogenesis mechanism studied here. Particles gain a large mass, compared to the temperature, when being swept up in relativistically expanding bubbles. The subsequent CP violating decays source the baryon asymmetry. Right: A sketch of the Azatov and Vanvlasselaer type production mechanism as a source of the baryon asymmetry. A fraction of the $\phi$ quanta pair produce $\Delta+\Delta^{\ast}$ across the bubble wall, due to an interaction term, which decay in a CP violating way to source the baryon asymmetry.
}
\label{fig:sketch}
\end{figure*}

In this paper, we shall explore the possibility of using either the simple mass generation (from now ``mass gain" or MG mechanism), or the Azatov and Vanvlasselaer (from now ``AV-type" production), during bubble expansion for baryogenesis. The idea is simple: the out-of-equilibrium particles in the broken phase following wall crossing decay in a CP and $B-L$ violating way, generating the baryon asymmetry~\cite{Sakharov:1967dj,Weinberg:1979bt,Kolb:1979qa}. A sketch of the two mechanisms is shown in Fig.~\ref{fig:sketch}. Although all the ingredients for the mass gain mechanism have been known for a long time it seems --- to the best of our knowledge --- this possibility has so far been missed in the literature. 

Conceptually, the MG mechanism seems to us the simplest way a phase transition can lead to the production of a baryon asymmetry. It is therefore well worth exploring. It has applications in a number of models, which can feature phase transitions around the scale of the decaying particles. This includes --- but is not limited to --- models of leptogenesis with or without gauge interactions for the decaying heavy states associated with opening up the Weinberg operator~\cite{Hambye:2003ka,Hambye:2003rt,Antusch:2004xy,Hambye:2005tk,Lavignac:2015gpa,Heeck:2016oda} (for an overview see~\cite{Hambye:2012fh}). Such leptogenesis models can also be associated with close-to-conformal dynamics~\cite{Marzo:2018nov,Brdar:2018vjq,Brdar:2018num,Brdar:2019iem,Brivio:2019hrj} such as occur in the ``neutrino option"~\cite{Brivio:2017dfq,Brivio:2018rzm,Brivio:2020aut}. In such models the calculation of the yield would need to be adapted to incorporate the effects discussed in the present paper, with details depending on the heavy neutrino mass scale and reheating temperature.

Explaining the matter asymmetry from a PT has previously been studied in great detail in the context of electroweak baryogenesis (EWBG)~\cite{Shaposhnikov:1987tw,Cohen:1990it}. If the wall velocity approaches the speed of light in EWBG, the yield of baryons approaches zero, due to suppressed particle diffusion back into the symmetric phase where the sphalerons are active~\cite{Cline:2020jre} (also see~\cite{No:2011fi,Caprini:2011uz}). This has consequences for the gravitational wave (GW) signal, as very strong PTs needed to produce a sizable signal typically also lead to ultra-relativistic walls, as shown via the Bodeker and Moore criterion~\cite{Bodeker:2009qy,Bodeker:2017cim}.

Moving away from EWBG, PTs have previously been considered in the context of leptogenesis in a number of studies. The Majorana mass, $M_{N}$, gained by the right handed neutrino of the type-I seesaw mechanism was considered in the context of phase transition dynamics during leptogenesis in~\cite{Shuve:2017jgj}. Relativistic walls which allow for $M_{N}/T_{n} \gg 1$ with no Boltzmann suppression, however, were not considered therein. Our mass gain mechanism could easily be applied to this scenario if the wall speed is relativistic. Let us also mention that the effect of a second order phase transition in low scale resonant leptogenesis was explored in~\cite{Pilaftsis:2008qt}, leptogenesis via a CP violating leptonic phase transition was examined in~\cite{Pascoli:2016gkf,Turner:2018mwh,Pascoli:2018cqk}, and diffusion type baryogenesis at a lepton number breaking phase transition was proposed in~\cite{Long:2017rdo}. But these are again, conceptually different, to the mechanisms considered here.

More exotic scenarios which instead actually rely on relativistic walls have also previously been considered. Heavy particle production from bubble wall collisions was studied in~\cite{Falkowski:2012fb} and was found to be (in)efficient for (in)elastic wall collisions. In the case of elastic wall collisions, it was subsequently shown that the bubble collisions can be used to explain the visible and dark matter densities~\cite{Katz:2016adq}. Whether the wall collisions are elastic or not depends on the shape of the potential~\cite{Falkowski:2012fb,Jinno:2019bxw}. (Another option, generating the baryon asymmetry via a beyond-the-standard-model instanton biased by the dynamics of a relativistic wall, was explored in~\cite{Cheung:2012im}.)

In this work, we instead assume inelastic wall collisions, so the heavy particles are produced during bubble expansion and not at collision (solely for simplicity, heavy particle production at wall collision would not necessarily invalidate our study). Inelastic collisions commonly occur in potentials with a large field distance between the false and true vacua at nucleation with modest barriers behind which the field has become temporarily stuck~\cite{Falkowski:2012fb,Jinno:2019bxw}. Such potentials produce thick walled bubbles at nucleation and are well known from close-to-conformal theories.

To be more specific, for the mass gain mechanism, we can have $M_{\Delta}=0$ in the symmetric phase and can eventually implement the phase transition in a classically scale invariant potential (as relevant in the aforementioned ``neutrino option"). Such models almost automatically result in supercooled phase transitions which are desirable for our baryogenesis mechanism. They are also favourable as they can return the desired supercooling without resulting in the field becoming permanently stuck in the false vacuum. For such close-to-conformal potentials, primordial black hole overproduction is also avoided~\cite{Lewicki:2019gmv}.

For the AV-type production, we instead need $M_{\Delta}\neq 0$ in the symmetric phase, which precludes an implementation of the above mechanism in a classically scale invariant theory. Qualitatively similar behaviour can, however, be mimicked by sufficiently flat single or multi-field potentials with appropriate thermal corrections, albeit in a more limited region of parameter space.

Historically, the production of heavy particles in a non-thermal fashion, which then source the baryon asymmetry through their decays, has been considered via inflaton decays~\cite{Dimopoulos:1987rk,Lazarides:1990huy,Asaka:1999yd,Hamaguchi:2001gw,Hamada:2018epb,Masoud:2019gxx}. Our MG mechanism, in particular, shares some conceptual similarities, but is not tied to the end of (slow roll) inflationary dynamics. Let us also mention that in supercooled confining phase transitions, the strong sector hadrons produced as fundamental quanta enter the bubble walls are also far from equilibrium, and that these could in principle also source the baryon asymmetry (although so far only considered for DM)~\cite{Baldes:2020kam,Baldes:2021aph}. In the current work, however, we will restrict ourselves to perturbative models which are under better computational control. 

As the MG and the AV-type mechanisms share basically all features --- apart from a few formulas --- we shall present them together below and make abundantly clear when differences occur, and also include superscripts ``MG" or ``AV" for mechanism dependent quantities or limits. We will first not specify a detailed potential for $\phi$, but instead investigate the required relations between $v_{\phi}$, $T_n$, $M_{\Delta}$, $\lambda$, the reheating temperature, $T_{\rm RH}$, and the couplings $y$ controlling the CP violation, in order for baryogenesis to be successful. We do so first in a rather model independent way, stating our assumptions at each step. We then illustrate the possible CP and $B-L$ violating decay modes by giving a simple example model for $\Delta$ and its interactions, which also allows us to study washout in greater detail. For the mechanism to work, we shall see that it is crucial that the reheating temperature $T_{\rm RH}$ is sufficiently below $M_{\Delta}$, so that $B-L$ violating collisions of the thermal bath do not washout the produced asymmetry. The expected gravitational wave signature is detailed and shown to be promising for testing via the Einstein Telescope, at least in some areas of the parameter space. Finally, we turn to some example potentials which reproduce the desired macroscopic (bulk) behaviour of the phase transition. For the simple example potentials studied here, however, the peak GW frequency ends up being too high for currently proposed interferometers.

We shall see that the current implementation works best at rather high scales (far above the electroweak one). We wish to emphasise, however, that simple extensions left for future exploration could easily allow one to lower the scale. To give an example, one may consider resonant enhancement of the CP violation. This has been studied extensively in the literature in the context of leptogenesis, as it allows one to overcome restrictions on the amount of CP violation for low right handed neutrino mass scales~\cite{Hambye:2001eu,Davidson:2002qv}. Similar conclusions hold for type-II and type-III leptogenesis in which the decaying particles carry gauge charges~\cite{Hambye:2012fh}. For a first pass, however, we will consider the CP violation in our model only in the hierarchical limit, and also limit ourselves to two-body decays for our heavy states.

\section{The phase transition and yield}

The temperature evolution of the Euclidean action, $S$, governing the bubble nucleation rate density, $\Gamma_{\rm Bub.} \sim T^{4}e^{-S}$, allows the universe to avoid getting stuck in the metastable state even if it is vacuum dominated (e.g. through a decreasing thermal barrier in the effective potential). This leads to cosmologically significant bubble nucleation at some $T_{n}$, conventionally defined when $\Gamma_{\rm Bub.} = H^{4}$, where $H$ is the Hubble rate. For most transitions bubble percolation occurs at roughly the same temperature, $T_{p} \approx T_{n}$, which we assume holds here. Overall, there are then two distinct possibilities before nucleation:
\begin{enumerate}[(i)]
\item  If the temperature drops far enough, $T_n < T_{\rm infl}$, the universe reaches a vacuum dominated (inflationary) epoch. Here $T_{\rm infl}$ is defined by
	\begin{equation}
 	\frac{g_{\ast}\pi^{2}}{30}T_{\rm infl}^{4} = \Lambda_{\rm vac} \equiv c_{\rm vac} v_{\phi}^{4}
	\end{equation}
where $c_{\rm vac}$ parametrizes the vacuum energy difference. Up to possible differences in $g_{\ast}$ between the two phases, which we assume small, $T_{\rm RH} \simeq T_{\rm infl}$. Note we also assume for now the $\phi$ condensate decays rapidly following the phase transition, meaning an early matter dominated epoch is avoided. This assumption can of course be checked once a concrete model is written down so that the $\phi$ decay rate can be evaluated. 

\item If instead, $T_n > T_{\rm infl}$, then reheating does not increase the temperature much, and we take $T_{\rm RH} \simeq T_{n}$. 
\end{enumerate}
As we shall see below, the production mechanism requires $T_n$ sufficiently below $v_{\phi}$, but a vacuum dominated phase is not mandatory. 

\subsection{Mass gain mechanism}
In the simple mass gain mechanism, we have effectively a massless $\Delta$ prior to the phase transition. Asymmetry generation in the thermal bath is suppressed due to the gauge interactions and the effectively massless $\Delta$. Through their various interactions, the $\Delta$ quanta are in thermal equilibrium with the bath prior to the phase transition and
	\begin{equation}
	 Y_{\Delta}^{\rm MG} \equiv \frac{n_{\Delta}^{\rm eq}}{s_e} =  \frac{ 45\zeta(3)g_{\Delta} }{ 2\pi^{4}g_{\ast} }
	\end{equation}
is their number density normalised to entropy, where $g_{\Delta}$ are the degrees of freedom of $\Delta$. Provided that $\gamma_{\rm w} > M_{\Delta}/T_{n}$ so the particles can enter the bubble, we have the $Y_{\Delta}$ maintained across the bubble wall. The $\Delta$ are then out-of-equilibrium, massive, and can decay in a CP and $B-L$ violating way. This gives a baryonic yield
	\begin{align}
	\frac{Y_{B}^{\rm MG}}{Y_{B}^{\rm Obs.}} & = \epsilon_{\Delta} \kappa_{\rm Sph.} \frac{  Y_{\Delta}^{\rm MG} }{ Y_{B}^{\rm Obs.} } \left( \frac{T_{n}}{T_{\rm RH}} \right)^{3} \nonumber \\
	& \approx 2.3 \times 10^{5} g_{\Delta}  \left(  \frac{ 100 }{ g_{\ast} } \right) \left(  \frac{ \epsilon_{\Delta} }{ 1/16\pi } \right)   \left( \frac{T_{n}}{T_{\rm RH}} \right)^{3}
	\label{eq:YByield0}
	\end{align}
where $\epsilon_{\Delta}$ is the average baryon asymmetry produced in each $\Delta$ or $\Delta^{\ast}$ decay, $\kappa_{\rm Sph.} = 28/79$ is a sphaleron reprocessing factor, the $(T_n/T_{\rm RH})^{3}$ factor takes into account the entropy production from reheating following the PT. We have normalised to the observed asymmetry $Y_{B}^{\rm Obs.} \approx 0.86 \times 10^{-10}$~\cite{Aghanim:2018eyx,Fields:2019pfx}. In the above formula we assume the $\Delta$ decay before annihilation once they cross the wall and that washout processes which remove the asymmetry are inefficient. Both assumptions will be critically evaluated later in the paper. From the above we see that the observed asymmetry can be explained provided the CP violation is not too suppressed and there is not too much entropy production following the phase transition. 

\subsection{AV-type production}

We now turn to the alternative scenario. Let us first remark, that in the AV-type scenario we will be interested in parameter space in which the decay rate of $\Delta$ will be large compared to Hubble, $\Gamma_{\Delta} > H$, at temperature $T=M_{\Delta}$. The number-density-to-entropy ratio of $\Delta$ prior to the PT will be completely negligible, as $M_{\Delta} \gg T_{n}$. Similarly, if $\Delta$ is charged under QCD --- as in our model below --- CP conserving annihilations keep $Y_{\Delta}$ close to its equilibrium value, $Y_{\Delta}^{\rm eq}$, at $T=M_{\Delta}$. (Unless $M_{\Delta}$ is close to the Planck scale $M_{\rm Pl}$.) The source term, for the usual baryogenesis from decays scenario, in the Boltzmann equation is therefore also suppressed around $T=M_{\Delta}$. Furthermore, $B-L$ violating $2 \leftrightarrow 2$ washout interactions also suppress any generated asymmety. One would naively guess the thermally produced asymmetry is therefore completely negligible.

Quantitatively, the picture is not so simple, as is known from analagous leptogenesis studies in which the decaying particles also carry gauge interactions~\cite{Hambye:2012fh}. Even if QCD annihilations are rapid, the initial state $\Delta+\Delta^{\ast}$ is doubly-Boltzmann suppressed, which leads to an eventual freezeout of the annihilations. An asymmetry can then be generated from decays of the frozen-out particles at late times.

Furthermore, our $\Delta$ particles will have multiple decay channels with final states not related by CP. Then --- as is known from type-II leptogenesis --- if one decay channel is faster than the annihilations, the latter will no longer suppress the asymmetry. Crucially, if the second decay channel is below $H$, washout is also suppressed. The price to pay is some reduction of the CP violation due to the differing branching ratios. Nevertheless, overall parameter space is opened up~\cite{Hambye:2005tk}.

The above discussion briefly summarizes why type-II and type-III leptogenesis is actually possible for masses, $M \gtrsim 10^{3}$ GeV  ($M \gtrsim 10^{10}$ GeV), of the heavy charged states in the resonant (hierarchical) regime of CP violation~\cite{Hambye:2003ka,Hambye:2003rt,Antusch:2004xy,Hambye:2005tk,Lavignac:2015gpa,Heeck:2016oda}. Although one would naively guess this is not possible due to the gauge interactions and the third Sakharov condition. We expect a broadly similar picture is also possible in our model.

\begin{figure}[t]
\begin{center}
\includegraphics[width=160pt]{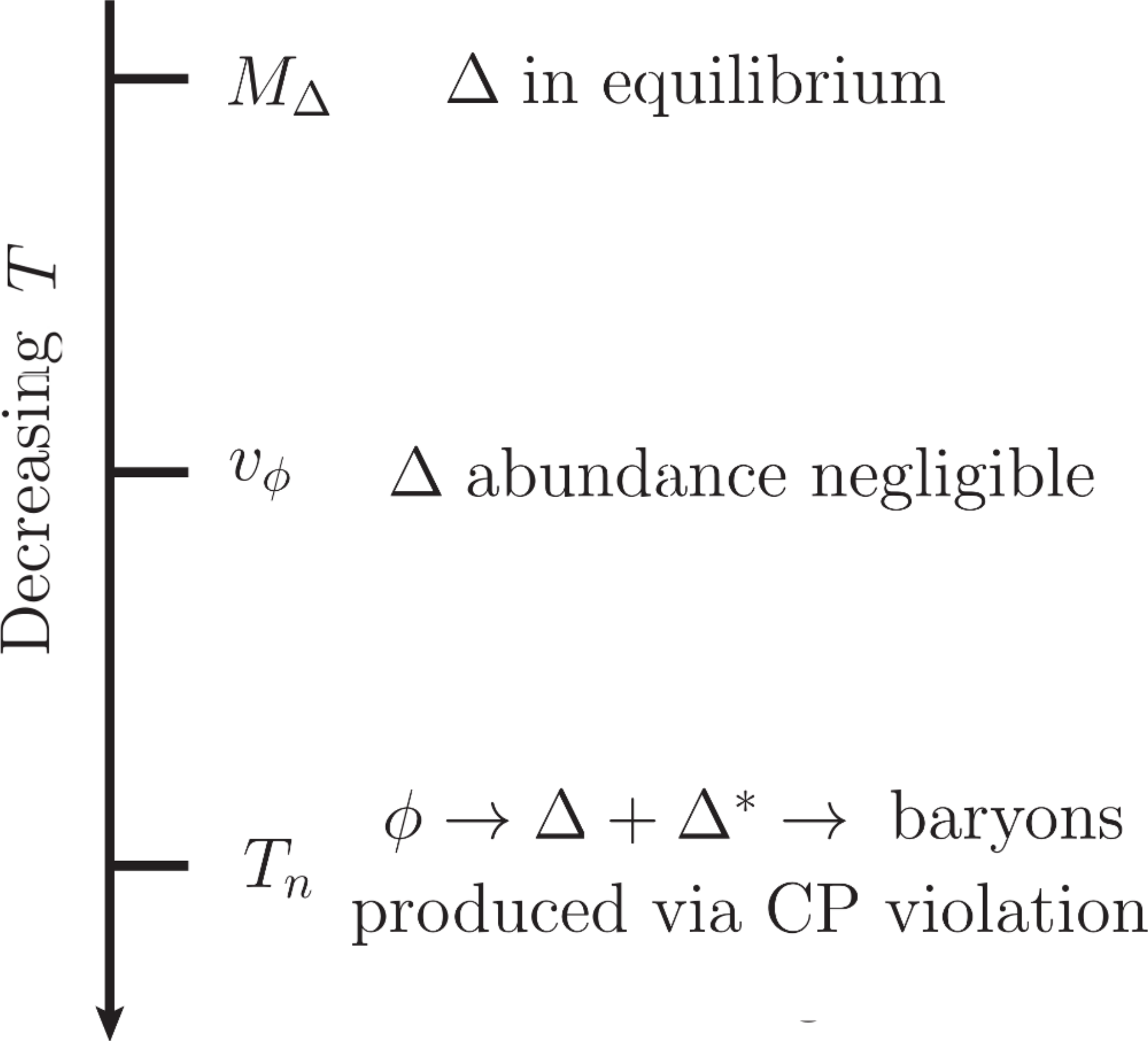}
\end{center}
\caption{Timeline of the mechanism for AV-type production. No asymmetry is produced until out-of-equilibrium $\Delta$ states are pair produced at $T_n$ when $\phi$ quanta encounter the walls. The $\Delta$ states then decay in a CP and $B-L$ violating way producing the baryon asymmetry of the universe.
}
\label{fig:diag}
\end{figure}

Precise determination of the usual asymmetry generation via standard thermal processes in the current model is left for future refinement. Here we solely focus on the baryon asymmetry created during the phase transition itself, which we assume gives the dominant contribution in determining the parameter regions of interest. If a non-negligible asymmetry is already generated around $T \simeq M_{\Delta}$, the PT can actually trigger a second burst of net baryon number generation. To obtain the observed asymmetry one can always tune relevant phases. A sketch of the timeline of AV-type production mechanism is shown in Fig.~\ref{fig:diag}.

Invoking the AV production, the yield of $\Delta$'s normalised to entropy following wall crossing is
	\begin{equation}
	Y_{\Delta}^{\rm AV} = P(\phi \to \Delta\Delta^*) Y_{\phi}^{\rm eq},
	\end{equation}
where $Y_{\phi} \approx 45\zeta(3)/(2\pi^{4}g_{\ast})$ is the equilibrium $\phi$ number density normalised to entropy with one degree of freedom. The $\Delta$'s will decay in a CP and $B-L$ violating way, giving a baryon asymmetry
	\begin{align}
	\frac{Y_{B}^{\rm AV}}{Y_{B}^{\rm Obs.}} & = \epsilon_{\Delta} \kappa_{\rm Sph.} \frac{  Y_{\Delta}^{\rm AV} }{ Y_{B}^{\rm Obs.} } \left( \frac{T_{n}}{T_{\rm RH}} \right)^{3} \nonumber \\
	& \approx 2.5 \times 10^2 \times g_{\Delta} \lambda^{2} \label{eq:YByield1} \\ 
	& \qquad \times \left(  \frac{ 100 }{ g_{\ast} } \right) \left(  \frac{ \epsilon_{\Delta} }{ 1/16\pi } \right) \left( \frac{T_{n}}{T_{\rm RH}} \right)^{3} \left( \frac{v_{\phi}}{M_{\Delta}} \right)^2, \nonumber
	\end{align} 
We see the baryon asymmetry can also be explained in this case, provided the CP violation is not too suppressed and the hierarchy in $M_{\Delta} > v_{\phi}$ does not become too severe, and there is not too much entropy production. We will show below that this is indeed consistently possible with other requirements of the model. The above formula again assumes the $\Delta$ decay before annihilation when crossing the wall and that washout processes are inefficient.

\subsection{Wall dynamics}

We now derive the conditions required to ensure we have sufficiently relativistic walls for our baryogenesis mechanisms. At leading order, the bubble wall feels a pressure from particles crossing the wall and gaining a mass. This is limited from above by the pressure calculated by Bodeker and Moore for ultra-relativistic walls~\cite{Bodeker:2009qy}
	\begin{equation}
	\mathcal{P}_{\rm LO} = \sum_{a} \Delta(m_{a}^{2})\int \frac{ d^{3}pf_{a}^{\rm eq} }{(2\pi^{3})2E_a}  \equiv  g_{a}\frac{ v_{\phi}^{2}T_{n}^{2} }{24} ,
	\label{eq:press}
	\end{equation}
where the sum runs over the degrees of freedom coupled to $\phi$, $f_{a}^{\rm eq}$ is their equilibrium number density in the symmetric phase, and $g_{a}$ parametrizes the effective number of light degrees of freedom coupled to $\phi$ convoluted with their change in mass.\footnote{Focussing on the $\Delta$ degrees of freedom in the MG mechanism, a similar estimate of the pressure (within $\approx 50\%$) results from simply taking $\mathcal{P}_{\rm LO} \approx n_{\Delta} \gamma_{\Delta}M_{\Delta} \approx g_{\Delta} \zeta(3) T_{n}^{2}M_{\Delta}^{2}/2\pi^2$, which shows the consistency of our approximation. Still, we use the Bodeker-Moore result as it is more precise.}
(If a gauge boson gains a mass during the phase transition, there is an additional important next to leading order pressure which grows with $\gamma$~\cite{Bodeker:2017cim}, although we do not consider this possibility further here.)
Note our assumption of a light $\phi$ and an unsuppressed $n_{\phi}$ we made in deriving $Y_{B}$ in Eq.~\eqref{eq:YByield1} above, is consistent with the induced thermal mass squared, $g_{a}T^{2}/24$, unless $g_{a} \gtrsim 200$. The $\phi$ mass of course also receives model dependent zero temperature contributions.

In order for the wall to reach a sufficiently high $\gamma_{\rm w}$, the vacuum expansion pressure, $c_{\rm vac}v_{\phi}^{4}$, should overcome $\mathcal{P}_{\rm LO}$. We therefore require 
	\begin{equation}
	\frac{T_n}{v_{\phi}} < \sqrt{\frac{ 24 c_{\rm vac} }{ g_{a} }} \approx 0.5  \times \sqrt{\left( \frac{ c_{\rm vac} }{ 0.1 } \right) \left( \frac{ 10 }{ g_a } \right)}.
	\end{equation}
Provided this is satisfied, the Lorentz factor at collision is given by the ratio of the bubble size at collision, $\sim M_{\rm Pl}/(\beta_H g_{\ast}^{1/2} T_{\rm RH}^{2})$, to that at nucleation, $\sim 1/T_n$. Here $\beta_{H} \equiv -TdS/dT|_{\rm nuc}$ controls the typical number of bubbles per Hubble volume, $N \sim \beta_{H}^{3}$. For strong phase transitions one typically finds $\beta_{H} \sim 10-1000$.

We have assumed above a relativistic wall with $\gamma_{\rm w} > M_{\Delta}/T_{n}$ in the mass gain mechanism, in order for the $\Delta$ quanta to be able to cross the wall rather than be reflected. This gives a constraint
	\begin{align}
	\frac{M_{\Delta}^{\rm MG}}{T_{n}} & < \frac{ M_{\rm Pl} T_{n} }{ \beta_{H} g_{\ast}^{1/2} T_{\rm RH}^{2} }  \\
	 &  \approx 10^{4} \left( \frac{ g_{\ast} }{ 100 } \right)^{\frac{1}{2}} \left( \frac{ 10} {\beta_{H} } \right)  \left( \frac{ T_{n} } {10 T_{\rm RH} } \right) \left( \frac{ 10^{12} \; \mathrm{GeV} }{ T_{\rm RH} } \right). \nonumber
	\end{align}
For the AV-type production we have instead assumed $\gamma_{\rm w} > M_{\Delta}^{2}/(T_{n}v_{\phi})$. We therefore require 
	\begin{align} 
	\frac{M_{\Delta}^{\rm AV}}{T_{n}} & < \sqrt{ \frac{ M_{\rm Pl} v_{\phi} }{ \beta_{H} g_{\ast}^{1/2} T_{\rm RH}^{2} } }  \\ & \approx 10^{3}  \left( \frac{ g_{\ast} }{ 100 } \right)^{\frac{1}{4}} \sqrt{ \left( \frac{ 10} {\beta_{H} } \right)  \left( \frac{ v_{\phi} } {10 T_{\rm RH} } \right) \left( \frac{ 10^{12} \; \mathrm{GeV} }{ T_{\rm RH} } \right) }. \nonumber
	\end{align}
Finally note that once the $\Delta$ particles are being produced via the AV-mechanism, there will be an additional retarding pressure on the bubbles, captured by $g_{a} \to g_{a} + P(\phi \to \Delta\Delta^*)$ in Eq.~\eqref{eq:press}. For finely balanced parameter points, this may prevent true runaway behaviour, and result in the energy density being carried in the plasma rather than in the scalar field configuration~\cite{Vanvlasselaer:2020niz}. This has possible consequences for the shape and amplitude of the gravitational wave spectrum~\cite{Caprini:2019egz}. However, it does not affect the baryognesis mechanism proposed here.

\section{The B and CP violation}
Having discussed the general properties required for the phase transition and mass spectrum, we now turn to illustrating a possible microphysical picture for the $\Delta$ decays. We introduce $i=1,2$ generations of $\Delta_{i}$, with $M_{\Delta 2} > M_{\Delta 1}$. We take these to now transform as $\Delta_i \sim (3,1,2/3)$ under the SM gauge group $SU(3)_{c} \times SU(2)_L \times U(1)_{Y}$ (with convention for the usual electromagnetic charge $Q_{\rm EM} = Y$ for an $SU(2)_L$ singlet). We introduce the following Yukawa interactions allowed by the gauge symmetries,\footnote{More generally one may consider any choice for the transformation properties of $\Delta$ which lead to the same form of diquark+leptoquark couplings and therfore break $B$. There are four possibilities in total, see e.g.~\cite{Baldes:2011mh}. All three other possibilities, however, conserve $B-L$ and therefore do not lead to asymmetry formation above $T_{\rm EW}\sim 100$ GeV, i.e.~when the sphalerons are active. The choice of Eq.~\eqref{eq:couplings} is special because of the gauge singlet fermion, which may hide the negative $B-L$ charge in the dark/singlet neutrino sector until below $T_{\rm EW}$, and therfore allow for baryogenesis. Introducing explicit $B-L$ violation is also possible by introducing differently charged $\Delta$'s and interactions between them, again see~\cite{Baldes:2011mh} and also~\cite{Bowes:1996ew}. A leptogenesis style realisation is briefly discussed in App.~\ref{sec:LGEN}.}
	\begin{equation}
	\mathcal{L} \supset y_{di}\Delta_{i}\overline{d_{R}^{c}}d_{R}' + y_{ui}\Delta_{i}\overline{N_{R}}u_{R}^{c} + \mathrm{H.c.},
	\label{eq:couplings}
	\end{equation}
where colour and flavour indices are suppressed. Note $y_{di}$ is antisymmetric in flavour, which can be seen by using $\overline{\psi^{c}}\chi = \overline{\chi^{c}}\psi$ and the antisymmetric property of the colour indices. Here $N$ is a gauge singlet fermion and we assume $M_{\Delta} \gg M_{N}$ to avoid complications with phase space effects. (We shall return to the possibilities for $N$ below.) If the above couplings are complex the interactions will violate CP. Some --- but not all --- of the phases can be removed through field rephasings. To be more precise with some examples:
	\begin{itemize}
	\item In the minimal case, there is one copy of $N_{R}$, only one down type flavour combination and one up type quark couples to the two $\Delta_{i}$. There are 4 couplings and 1 physical phase after field rephasings.
	\item For one copy of the $N_R$, we find 12 (6) independent couplings with 5 (2) physical phases if couplings to the first generation quarks are (not) included and all other couplings are present.
	\item For three copies of the $N_R$, we find 24 (14) independent couplings with 15 (8) physical phases if couplings to the first generation quarks are (not) included and all other couplings are present.
	\end{itemize}
In the above analysis, we have allowed rephasing of the $N_{R}$. If Majorana, this reintroduces a phase in the $N_{R}$ mass term, $M_{N}$, which would then need to be included in calculations of the CP violation. This is not expected to play a large role, as we anyway take the limit $M_{N} \ll M_{\Delta}$ for our calculation of the CP violation below, in order to avoid kinematic suppression. Suffice to say, that the above couplings can easily lead to CP violating decays.

\begin{figure}[t]
\begin{center}
\includegraphics[width=260pt]{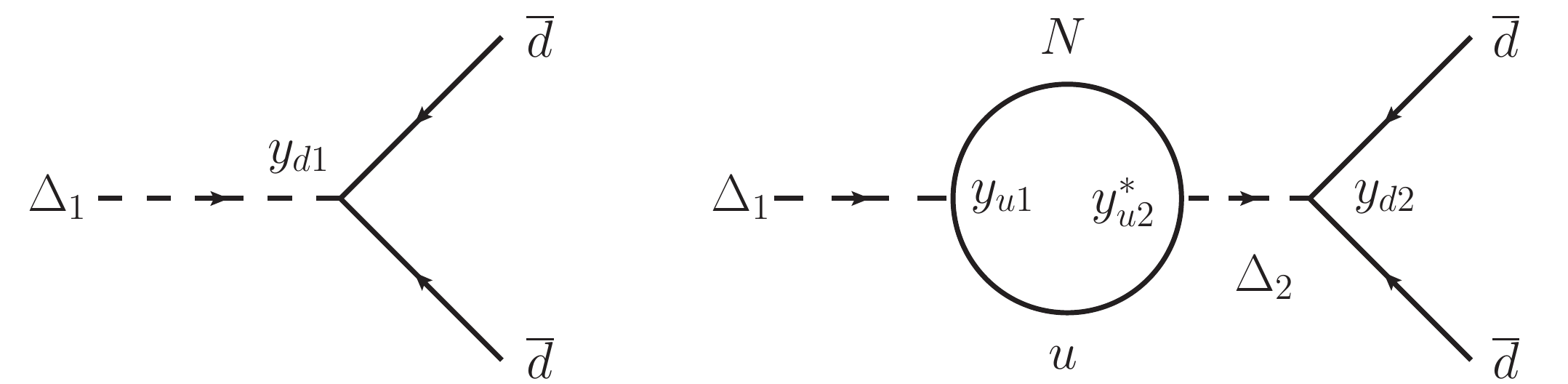}
\end{center}
\caption{Example of the diagrams which interfere and lead to CP violation. The intermediate loop particles are kinematically able to go on-shell. 
}
\label{fig:inter}
\end{figure}

Similar couplings have previously been studied in the context of baryogenesis in a number of papers, but typically with a focus on generating the asymmetry via decays or scattering of the $N$ after the $\Delta_{i}$ have been integrated out~\cite{Katz:2016adq,Claudson:1983js,Dimopoulos:1987rk,Cheung:2013hza,Baldes:2014rda,Davoudiasl:2015jja,Dev:2015uca,Grojean:2018fus}.

The above couplings lead to tree level decay rates
	\begin{align}
	\Gamma(\Delta_{i} \to \overline{d_{R}}\overline{d_{R}'}) \approx \frac{ |y_{di}|^2 }{8\pi}M_{\Delta i}, \\ 
	\Gamma(\Delta_{i} \to Nu_{R}) \approx \frac{ |y_{ui}|^2 }{16\pi}M_{\Delta i}.
	\end{align}
Here, we have summed over the final state colours (for the first decay) but leave the summation over flavours implicit. Interference between tree and loop level diagrams, as illustrated in Fig.~\ref{fig:inter}, leads to CP violation in the decays. Focusing on $\Delta_{1}$, we parametrize the CP violation as
	\begin{align}
	\Gamma(\Delta_{1} \to \overline{d_{R}}\overline{d_{R}'}) \equiv \Gamma_{1d}(1+\epsilon_{d}), \\
	\Gamma(\Delta_{1}^{\ast} \to d_{R}d_{R}') \equiv \Gamma_{1d}(1-\epsilon_{d}), \\
	\Gamma(\Delta_{1} \to Nu_{R}) \equiv \Gamma_{1u}(1+\epsilon_{u}), \\
	\Gamma(\Delta_{1}^{\ast} \to \overline{N}\overline{u_{R}}) \equiv \Gamma_{1u}(1-\epsilon_{u}).
	\end{align}
The total decay rates of $\Delta_{1}$ and $\Delta_{1}^{\ast}$ must be equal ($\equiv \Gamma_{\Delta}$) so $\epsilon_{d}\Gamma_{1d} = -\epsilon_u\Gamma_{1u}$. The average baryon asymmetry produced in each $\Delta_1$ or $\Delta_{1}^{\ast}$ decay is then $\epsilon_{\Delta} = \epsilon_{u}\Gamma_{1u}/\Gamma_{\Delta}$, which can be substituted into either Eq.~\eqref{eq:YByield0} or \eqref{eq:YByield1}. Using the Cutkosky rules~\cite{Cutkosky:1960sp,Kolb:1990vq,Peskin:1995ev,Millar:2014afa} to extract the imaginary part of the loop in Fig.~\ref{fig:inter}, we find
	\begin{align}
	\epsilon_{\Delta} & = \frac{1}{2\pi}\frac{\mathrm{Im}(y_{d1}^{\ast}y_{u1}y_{u2}^{\ast}y_{d2})}{|y_{u1}|^{2}+2|y_{d1}|^{2}}\frac{M_{\Delta 1}^{2}}{M_{\Delta 2}^{2} -M_{\Delta 1}^{2}} \nonumber \\ & \sim \frac{ \mathrm{Im}[y^{2}] }{ 6\pi } \left( \frac{ M_{\Delta 1} }{ M_{\Delta 2} } \right)^{2},
	\end{align}
where in the second line we assume no major hierarchies in the Yukawa couplings, $y$, and generic $\mathcal{O}(1)$ phases. Although $y_{di}$ is antisymmetric in flavour, there is importantly no relative minus sign when summing over the flavours in the numerator above as both $y_{d1}^{\ast}$ and $y_{d2}$ change sign, and so $\epsilon_{\Delta} \neq 0$ is indeed possible.

Substituting the CP violation into Eq.~\eqref{eq:YByield0} we find for the mass gain option
	\begin{equation}
	\frac{Y_{B}^{\rm MG}}{Y_{B}^{\rm Obs.}} \sim 7 \times 10^{5} g_{\Delta} \mathrm{Im}[y^{2}] \left(  \frac{ 100 }{ g_{\ast} } \right) \frac{   T_{n}^{3} M_{\Delta 1}^2 }{ T_{\rm RH}^{3} M_{\Delta 2}^{2} }.
\label{eq:YByield2a}
	\end{equation}
Instead substituting into Eq.~\eqref{eq:YByield1} we find for the AV-type option
	\begin{align}
	\frac{Y_{B}^{\rm AV}}{Y_{B}^{\rm Obs.}} \sim 7.5 \times 10^{2} g_{\Delta} \lambda^{2} \mathrm{Im}[y^{2}] \left(  \frac{ 100 }{ g_{\ast} } \right) \frac{   T_{n}^{3} v_{\phi}^2 }{ T_{\rm RH}^{3} M_{\Delta 2}^{2} }.
\label{eq:YByield2}
	\end{align}
Note the appearance of the heavy mass $M_{\Delta 2}$ and also the CP violating phases. Here, we do not consider the resonantly enhanced regime $M_{\Delta 2} - M_{\Delta 1} \lesssim M_{\Delta 1}$, which also requires more careful treatment of the propogators in the loop level diagrams for mass splittings below $\mathcal{O}(\Gamma_{\Delta})$.  

In our analysis of the previous section, we assumed $\phi$ decays rapidly following the phase transition, in order to avoid a matter dominated epoch before reheating and hence additional dilution of the baryon asymmetry. The $\phi$ decay channels are, in general, model dependent. In the current realisation, however, there are automatically decay channels to gluons and weak hypercharge gauge bosons at loop level through a $\Delta$ triangle diagram. The decay rates are
	\begin{align}
	& \Gamma(\phi \to gg) \sim \frac{\lambda^2 \alpha_s^2 v^2 m_\phi^3}{32 \pi M_\Delta^4}, \\
	& \Gamma(\phi \to YY) \sim \frac{\lambda^2 \alpha_Y^2 Q^4 v^2 m_\phi^3 }{32 \pi M_\Delta^4},
	\end{align}
where $\alpha_{s} \equiv g_{s}^{2}/4\pi$, and  $\alpha_{Y} \equiv g_{Y}^{2}/4\pi$ are the fine-structure constants. This automatically ensures the rapid decay of the $\phi$ condensate compared to $H$, provided 
	\begin{align}
	\frac{m_\phi}{M_\Delta} & \gtrsim 
	70 \left[\left(\frac{\alpha_s}{0.03}\right)^2+ 0.02 \left(\frac{\alpha_Y}{0.01}
	\right)^2\right]^{-1/3}
	\frac{  c_\text{vac}^{1/6} M_\Delta^{1/3}}{\lambda^{2/3} M_\text{Pl}^{1/3}} \nonumber \\
	& \approx 0.3 \left[\left(\frac{\alpha_s}{0.03}\right)^2+ 0.02 \left(\frac{\alpha_Y}{0.01}
	\right)^2\right]^{-1/3}
	 \\ 
	& \qquad \qquad \times \frac{c_\text{vac}^{1/6}}{\lambda^{2/3}}\left( \frac{M_{\Delta}}{10^{12} \; \mathrm{GeV}} \right)^{1/3} \nonumber
	\end{align}
For smaller $m_{\phi}$, however, it is easy to envisage additional decay modes for $\phi$, which do not pose any model building challenge. We shall assume these are present if required.

\section{Avoiding washout}
\label{sec:avoidingwashout}
\subsection{At $T_{n}$: just after wall crossing but before reheating}
\label{sec:avoidingwashoutA}
We first need to check that the $\Delta$'s, when they have just entered or been produced by the bubble wall, do indeed undergo CP violating decay before they annihilate with each other. We remind the reader, that in the plasma frame, the typical Lorentz factor of the produced $\Delta$ is $\gamma_{\Delta} = M_{\Delta}/T_{n}$, which will come in handy below.

\subsubsection{Mass gain mechanism} After crossing the wall, the $\Delta$ density in their own ``gas" rest frame is
	\begin{equation}
	n_{\Delta}^{\rm MG} = \frac{ n_{\Delta}^{\rm eq}(M_{\Delta} \simeq 0) }{ \gamma_{\Delta}(1-v_{\Delta}) } \approx  2 \left(\frac{M_{\Delta}}{T_{n}}\right) n_{\Delta}^{\rm eq}(M_{\Delta} \simeq 0),
	\end{equation}
where $n_{\Delta}^{\rm eq}(M_{\Delta}\simeq 0) \simeq g_{\Delta}\zeta(3)T_{n}^{3}/2\pi^2$ is the plasma frame equilibrium density. The factor $1/(1-v_{\Delta})$ comes from a squeezing together of the $\Delta$'s in one spatial dimension, due to their finite velocity, $v_{\Delta} = \sqrt{1-1/\gamma_{\Delta}^{2}} \simeq 1 - T_{n}^{2}/2M_{\Delta}^2$, with respect to the plasma frame once they cross the wall. (The final $1/\gamma_{\Delta}$ factor comes from transforming from the plasma to the $\Delta$ frame.)

The $\Delta$'s have an irreducible annihilation back into $\phi$~\cite{Shuve:2017jgj}, whose mass is naturally $m_{\phi} \lesssim v_{\phi}$ for strong phase transitions. As our $\Delta$ carries colour and weak hypercharge, annihilations into gauge bosons can, of course, also occur. The annihilation cross sections in the non-relativistic limit are given by
	\begin{align}
	v_{\rm rel}\sigma(\Delta\Delta^{\ast} \to \phi\phi) &  \simeq \frac{ \pi \alpha_{\phi}^2 }{ 9M_{\Delta}^{2} } \left(1 - \frac{ 8\pi \alpha_{\phi} v_{\phi}^{2} }{M_{\Delta}^{2}} \right)^{2} S_0^{[1]}, \\ 
	v_{\rm rel}\sigma(\Delta\Delta^{\ast} \to gg) & \simeq \frac{14}{27} \frac{\pi \alpha_{s}^{2} }{ M_{\Delta}^{2} }  \left( \frac{2}{7} S_0^{[1]} + \frac{5}{7}S_0^{[8]}\right), \\
	v_{\rm rel}\sigma(\Delta\Delta^{\ast} \to YY) & \simeq \frac{8}{81} \frac{\pi \alpha_{Y}^{2} }{ M_{\Delta}^{2} } S_0^{[1]},
	\end{align}
for the annihilation into scalars, gluons and weak hypercharge gauge bosons respectively. Here $\alpha_{\phi} \equiv \lambda/8\pi$ is a parametrization chosen for convenience below (note the single power of $\lambda$), and $S_0^{[1]}$ and $S_0^{[8]}$ are Sommerfeld enhancment (SE)~\cite{Sommerfeld:1931,Sakharov:1948plh,March-Russell:2008lng} factors discussed further below, for details in approximately the current context see~\cite{Harz:2017dlj}. (The determination of the long distance SE factor factorizes with the hard scattering part of the annihilation cross section.) For the annhilation into scalars we have assumed a negligible $\phi$ self-quartic which is well justified in the context of a strong phase transition. Annihilations via the gauge bosons into SM fermions are p-wave and therefore suppressed. 

The relative velocity between the $\Delta$'s in their gas frame is $v_{\rm rel}^{\Delta} \sim T_{n}/M_{\Delta}$. This low $v_{\rm rel}$ can allow for a SE of the above annihilations via long range light boson exchange. The potential seen by the incoming non-relativistic particles depends on the colour representation of the incoming $\Delta+\Delta^{\ast}$ states. The potential can be written as
	\begin{equation}
	V(r) = -\frac{\alpha_{g}}{r}e^{-m_{g}r}-\frac{\alpha_{Y}Q_{Y}^2}{r}e^{-m_{Y}r}-\frac{\alpha_{\phi}}{r}e^{-m_{\phi}r},
	\label{eq:potNR}
	\end{equation}
where $m_{g} = \sqrt{8\pi\alpha_{s}}T$ is the thermal mass of the gluon, $m_{Y} = \sqrt{22\pi\alpha_{Y}/3}T$ is the thermal mass of the $Y$-boson, $\alpha_{g}=4\alpha_s/3$ ($-\alpha_s/6$) for the attractive (repulsive) gluon mediated colour singlet (octet) state~\cite{Harz:2017dlj}, and $Q_{Y} = 2/3$. The SE factors are functions of
	\begin{equation}
	\zeta_{g,Y,\phi} \equiv \frac{\alpha_{g,Y,\phi}}{v_{\rm rel}}, \qquad d_{g,Y,\phi} \equiv \frac{\alpha_{g,Y,\phi}M_{\Delta}}{2m_{g,Y,\phi}},
	\end{equation} 
and are given by $S_{0} \equiv |\phi_{\bf k}(0)|^{2}$. Here $\phi_{\bf k}$ is the solution to the Schr\"odinger equation
	\begin{equation}
	\left\{ \nabla_{z}^{2}+1+\frac{2}{z} \sum_{i=g,Y,\phi} \zeta_{i}Q_{i}^2 \mathrm{Exp}\left[ -\frac{ \zeta_{i}Q_{i}^2z}{d_{i}} \right] \right\}\phi_{\bf k} = 0,
	\end{equation} 
where ${\bf k}={\bf v}_{\rm rel}M_{\Delta}/2$ and ${\bf z}=k {\bf r}$, and $Q_{g,\phi} \equiv 1$ for compact notation. The boundary conditions are for an incoming plane wave and outgoing spherical wave at asymptotically large $r$. Numerically solving for the $S_0$ is left for future exploration. Here we estimate the possible size of the effect. Let us consider the Coulomb regime for all three bosonic potentials, $\alpha_{g,Y,\phi} M_{\Delta} \gg 1.68m_{g,Y,\phi}$. Then the SE factor becomes
	\begin{equation}
	S(\zeta_{\rm eff}) = \frac{2\pi\zeta_{\rm eff}}{1-e^{-2\pi\zeta_{\rm eff}}},
	\end{equation}
where $\zeta_{\rm eff} \equiv \sum_{i=g,Y,\phi}Q_{i}^{2}\alpha_i/v_{\rm rel}$. For $\zeta_{\rm eff} \gtrsim 0.5$, we have $S(\zeta_{\rm eff}) \simeq 2\pi\zeta_{\rm eff}$.

In order to be completely safe from $\Delta$ suppression, we require the total annihilation rate, $\Gamma_{\rm ann.} = n_{\Delta}^{\rm MG}\sigma v_{\rm rel}$, to be below the decay rate $\Gamma_{\Delta} \approx 3y^{2}M_{\Delta}/16\pi$. In order to gain some intuition, assume, for now, that the annihilation into scalars dominates. This can realistically be the case, as we require $\lambda \gtrsim 2$ for our mechanism to work, which translates into $\alpha_{\phi} \approx 0.08 \gtrsim \alpha_{s} \sim \alpha_{Y}$ for the high scales, $ \mu \gtrsim 10^{12}$ GeV, we will largely be interested in below. (The QCD channel dominates for $\mu \lesssim 10^{9}$ GeV. Whenever required, we estimate the gauge coupling strengths by the SM RGEs from~\cite{Buttazzo:2013uya}.) We also assume that $M_{\Delta} \simeq 0$ in the symmetric phase, so we have $M_{\Delta}^2 \simeq 4\pi\alpha_{\phi} v_{\phi}^{2}$, and $v_{\rm rel}\sigma(\Delta\Delta^{\ast} \to \phi\phi) \simeq \pi \alpha_{\phi}^{2}/(9M_{\Delta}^2) \times S_0^{[1]}$. Then if we are out of the SE regime, we require
	\begin{equation}
	y^{\rm MG} \gtrsim  \frac{\lambda}{\pi}  \sqrt{ \frac{ g_{\Delta}\zeta(3) }{ 108 }}  \frac{ T_n }{ M_{\Delta} }.
	\label{eq:deccond1}
	\end{equation}
If instead, we have a scalar mediated SE annhilation, we require
	\begin{equation}
	y^{\rm MG} \gtrsim \frac{\lambda^{3/2}}{\pi} \sqrt{ \frac{ g_{\Delta}\zeta(3) }{ 432 }} \sqrt{\frac{ T_n }{ M_{\Delta} }}.
	\label{eq:deccond2}	
	\end{equation}
Areas in which the above constraint is not met, assuming the SE Coulomb regime for the $\phi$ exchange, will be indicated on our summary plots below. Note the above conditions are actually somewhat strict. Provided the $Y_{B}$ yield is high enough, we can actually live with some $\Delta \Delta^{\ast}$ annihilation, which effectively switches off when the number density reaches $n_{\Delta} = \Gamma_{\Delta}/\sigma v_{\rm rel}$ (or if $n_{\Delta}\sigma v_{\rm rel}$ drops below $H$, but which we do not consider further here). In the non-SE regime, if inequality~\eqref{eq:deccond1} is violated, we then have a suppression factor of the yield
	\begin{equation}
	Y_{B}^{\rm MG} \to \frac{108\pi^{2}}{g_{\Delta}\zeta(3)} \frac{y^{2}M_{\Delta}^2}{\lambda^{2}T_{n}^{2}} \times Y_{B}^{\rm MG}
	\end{equation}	
In the SE regime, if inequality~\eqref{eq:deccond2} is violated, we instead have a suppression factor of the yield given by
	\begin{equation}
	Y_{B}^{\rm MG} \to \frac{432\pi^{2}}{g_{\Delta}\zeta(3)} \frac{y^{2}M_{\Delta}}{\lambda^{3}T_{n}} \times Y_{B}^{\rm MG}.
	\end{equation}

Note for low decay rates, $\Gamma_{\Delta} < \gamma_{\Delta}\beta_{H}H$, the $\Delta$ particles may actually decay only once the bubbles have percolated and the particles have crossed the collision plane. The boosted particles from opposing bubbles will then encounter one another. Consider a $\Delta$ interacting with the population coming from the opposing bubble. The density seen --- compared to the gas rest frame density we have considered above --- is increased by $\gamma_{\Delta}^{2}$. But the Mandlestam variable $s$ is boosted by the same amount. As $\sigma v_{\rm rel} \propto 1/s$, the annihilation rate, $\Gamma_{\rm ann.} = n\sigma v_{\rm rel}$, is not significantly increased. Furthermore, in this case, there is no SE as $v_{\rm rel} \simeq 1$. We will indicate the regions of parameter space in which the decay occurs in the opposing bubble in our plots below.

In order to keep our calculations tractable, we have assumed we are always in the Coulomb regime. Outside of this, narrow parametric resonances in the SE appear, corresponding to quasi-bound states with zero binding energy. In order to accurately take these effects into account, together with the interplay of the various terms in Eq.~\eqref{eq:potNR}, we would need to solve the Schr\"odinger equation and properly average over the velocity distributions. As the resonances are very narrow and occur only at very specific points, from which one can presumably escape by tiny changes in the parameters, we leave the exploration of such effects in the context of the current baryogenesis mechanism to future work.

Another interesting question is the formation of physical $\Delta-\Delta^{*}$ bound states via radiative emission of one of the mediating particles. The bound state formation via real scalar emission is suppressed~\cite{Oncala:2018bvl}. But the emission of one of the gauge bosons can lead to significant radiative capture cross sections. The capture via $Y$ emission is typically an $\mathcal{O}(1)$ value larger than the associated annihilation process in the large $\alpha_{Y}/v_{\rm rel}$ limit. The capture via $g$ emission can be an order of magnitude larger than the associated annihilation process --- albeit over a narrower range of $\alpha_{s}/v_{\rm rel}$ --- due to exponential suppression at low $v_{\rm rel}$ from the repulsive octet state. (To form a bound state via gluon emission, the final state must be in the attractive singlet representation, which means the initial state is in the octet.) For example, this has been explored in the DM context~\cite{Kim:2016zyy,Kim:2016kxt,Mitridate:2017izz,Keung:2017kot,Biondini:2017ufr,Braaten:2017kci,Braaten:2017dwq,Biondini:2018pwp,Harz:2018csl}, and of course also in QCD, e.g.~see~\cite{Pineda:1997bj,Beneke:1999zr,Brambilla:2011sg}. A detailed calculation requires the inclusion of all mediating particles participating in the interaction and is beyond the scope of the work. (The effect of the SM Higgs in the context of bound state formation of colour triplets has recently been studied in detail by Harz and Petraki~\cite{Harz:2019rro}.)

Instead, we can provide a simple criterion to see whether the bound state formation and subsequent decay (into $\phi\phi$, $gg$, $YY$) can be efficient in depleting the $\Delta$'s and suppressing $Y_{B}$. Let us again assume we are in the $\alpha_{\phi} \gtrsim \alpha_{s},\alpha_{Y}$ regime. Then the decay rate of the bound state scales as
	\begin{equation}
	\Gamma([\Delta\Delta^{\ast}]_{\rm Bound} \to \phi\phi) \sim \alpha_{\phi}^{5}M_{\Delta}.
	\end{equation}
It is a sufficient condition, in order to be safe from $\Delta$ suppression via this process, for the $\Delta$ constituents of the bound states to undergo $B$ violating decay more quickly than the bound state itself undergoes $B$ conserving decay. We are therefore safe if
	\begin{equation}
	y^{\rm MG} \gtrsim \lambda^{5/2} \times 10^{-3},
	\end{equation}
which is relatively easily satisfied for our baryogenesis mechanism. Note the necessary condition may actually be weaker. The reason is the bound states, in their frame, see a flux of highly energetic gluons coming from outside the bubbles, with $E_{g} \sim M_{\Delta}$ and density $n_{g} \sim T^{2}M_{\Delta}$. These may disassociate the bound states more quickly than they can decay, depending on the details of the out-of-equilibrium ionisation process. Careful evaluation is left for future work, e.g.~in the case of resonantly enhanced CP violation, in which case the Yukawa couplings may be much smaller than what is typically considered here. 

Finally, there are also hard scattering processes --- such as $\Delta+d \to \overline{d} \to g+\overline{d}$ with the thermal bath particles --- which lead to the effective removal of $\Delta$. These have a rate $\sim \gamma_{\Delta}\alpha_{\rm s} y^{2}T_n^{3}/M_{\Delta}^{2}$. Such scatterings are suppressed compared to the decay rate provided $g_{s} \lesssim M_{\Delta}/T_{n}$, which is easily satisfied as the right hand side (RHS) will realistically always be at least $\sim 10$.

\subsubsection{AV-type production}
For the AV-type production, the $\Delta$ density in their own gas rest frame is
	\begin{align}
	n_{\Delta}^{\rm AV} \approx \frac{n_{\phi}}{\gamma_{\Delta}(1-v_{\Delta}) } \times P(\phi \to \Delta\Delta^*),
	\end{align}
where $n_{\phi} \simeq g_{\phi}\zeta(3)T_{n}^{3}/2\pi^2$. We have $n_{\Delta}^{\rm AV} < n_{\Delta}^{\rm MG}$ and the annihilation rates are even more suppressed compared to the decay rate. For example, if we again assume the dominant annihilation channel is non-Sommerfeld enhanced $\Delta\Delta^{\ast}\to\phi\phi$ annihilation, we find a constraint 
	\begin{equation}
	y^{\rm AV} \gtrsim \lambda^2 \times \frac{ \sqrt{2 g_{\Delta}\zeta(3)} }{ 72\pi^{2}  } \frac{  v_{\phi} T_n }{ M_{\Delta}^{2} }.
	\end{equation}
Sommerfeld enhancement is expected to play less of a role in the AV-type production. This is because the produced particles have an enhanced velocity in the direction transverse to the wall. (Approximately flat in $k_{T}^{2}$ out to $k_{T} \sim M_{\Delta}$~\cite{Vanvlasselaer:2020niz,Azatov:2021ifm}.) We therefore have something closer to $v_{\rm rel} \sim 1$ on average in the AV-type production. Hence, long range effects such as SE and bound state formation are suppressed. And the decays dominate over annihilations more broadly, at least in the parameter space which can explain the observed baryon asymmetry.

\subsubsection{Energy and possible washout from decay products}

When the $\Delta$ enter or are produced inside the bubble, they carry a boost $\gamma_{\Delta} \approx M_{\Delta}/T_{\rm n}$ in the plasma frame. Let us estimate their kinetic thermalisation time with the plasma through gluon exchange with the thermal bath. This includes $g_{\mathrm{QCD}\ast}=79$ effective QCD degrees of freedom, with density $n_{\rm QCD} = g_{\mathrm{QCD}\ast}\zeta(3)T_{n}^{3}/2\pi^2$. Denote the energy transferred per single scattering as $\delta E$. The maximum energy transfer is given by a hard scattering $\epsilon_{\rm max} \approx \sqrt{E_{\Delta}T_n} \approx M_{\Delta}$. The minimum energy transfer is set by the gluon thermal mass $\epsilon_{\rm min} \approx m_{g}(T_{n})$. Then we estimate the energy transfer rate between the $\Delta$'s and the thermal bath, normalised to the initial energy, as~\cite{Baldes:2020kam}
	\begin{align}
	\frac{1}{E_{\Delta}}\frac{dE_{\Delta}}{d\tau} & \approx  \frac{n_{\rm QCD}}{E_{\Delta}} \int_{\epsilon_{\rm min}}^{\epsilon_{\rm max}} d( \delta E ) \delta E \frac{d\sigma}{d (\delta E)} \\
	 		     & \approx 	\frac{n_{\rm QCD}}{E_{\Delta}} \int_{-s}^{-m_g} dt \sqrt{-t} \frac{d\sigma}{dt} \\
			     & \approx \frac{4\pi \alpha_{s} n_{\rm QCD} }{ E_{\Delta}m_g(T_n)},
	\end{align} 
where $\tau$ is the time in the plasma frame, $t$ is the Mandelstam variable, and in the second line we have taken $v_{\rm rel}=1$, and assumed small $\delta E$ dominance so $t \simeq -(\delta E)^2$. In the last line we have approximated the cross section as $\sigma \sim 4\pi\alpha_{s}^2/(-t)$.

The thermalisation time is found by simply integrating the final line above with respect to $\tau$, setting the result to unity, and solving for what is now $\tau_{\rm th.}$. The kinetic thermalisation rate (i.e. the inverse thermalisation timescale), is then simply 
	\begin{equation}
	\Gamma_{\rm th.} = \frac{4\pi \alpha_{s} n_{\rm QCD} }{ E_{\Delta}m_g(T_n)} \approx \frac{ g_{\mathrm{QCD}\ast}\zeta(3) }{ \sqrt{2\pi^3} } \frac{ \alpha_{s}^{3/2}T_{n}^{3}}{M_{\Delta}^{2}}
	\label{eq:kineqdelta}
	\end{equation}
which can be easily compared to the $\Delta$ lifetime in the plasma frame, $\sim 3y^{2}T_{n}/16\pi$. Indeed for 
	\begin{equation}
	y \gtrsim \left( \frac{\alpha_s}{0.03} \right)^{3/4} \left( \frac{ g_{\mathrm{QCD}\ast} }{ 79 } \right)^{1/2} \frac{ T_{n} }{ M_{\Delta} },
	\label{eq:boostcond}
	\end{equation}   
the $\Delta$'s will decay with a majority of their initial boost preserved.\footnote{The above is a simplified analysis. In the MG mechanism, for $M_{\Delta}/T_{n} \gtrsim 10$, there is more energy in the $\Delta$ particles than in the thermal bath. Thus, there will be some reheating of the thermal bath due to the $\Delta$ interactions and decays, not just from the decay of the $\phi$ condensate. For the AV mechanism there is more energy in the $\Delta$ particles than in the thermal bath for $v_{\phi}/T_n \gtrsim \lambda\times 300$.} In the MG mechanism, $y$ can be somewhat below one (indeed we shall see $y \sim 0.01-0.1$ is preferred to avoid additional washout) while still explaining $Y_{B}^{\rm Obs.}$. So it is a question of the parameters chosen whether the $\Delta$'s decay while boosted or not. In the AV-type production, we instead need $y \gtrsim 1$, in order to produce enough baryons. There is also typically a larger hierarchy between $T_n$ and $M_{\Delta}$. So in this latter scenario, the $\Delta$'s will decay with their boost intact. We now discuss the possible washout effects of the boosted decay products. These are present once the asymmetry is first generated and will eventually return to kinetic equilibrium through their interactions. The question is whether any $B$ violating scatterings they undergo before returning to $T_{n}$ will lead to washout.

\underline{Case A: $\Delta$ decays with boost} 
We first consider the case in which inequality~\eqref{eq:boostcond} is satisfied. The decay products of the $\Delta$ have a typical energy $\sim M_{\Delta}^{2}/2T_{n}$ in the plasma frame. These can scatter with thermal bath particles of energy $T_{n}$. The centre-of-mass energy in such a collision is $\sim M_{\Delta}$ and so can resolve the $\Delta$ peak in the cross section, i.e. create another on-shell $\Delta$. As we shall see, off shell scatterings away from the resonance can also play a role, although typically subdominant.

In a hard scattering, the outgoing particle energies in the plasma frame, have energies approximately 1/2 the value of the incoming energy. There may therefore be a cascade of such interactions, with $N_{\rm steps} \approx 2 \, \mathrm{log}_2(M_{\Delta}/T_n) \sim (5-13)$ steps, if other processes in the plasma do not first cut off the $B$ violating hard scattering. The question we want to answer is whether such a sequence of interactions during the return to equilibrium will washout the asymmetry. 

Note, however, that even if not suppressed, the on-shell scatterings close to the $\Delta$ peak will also lead to CP violation of the same order as $\epsilon_{\Delta}$. As the initial state is highly out-of-equilibrium, these inverese decays will in effect simply reprocess the asymmetry to a numerically similar value. We shall see, however, that a danger arises because the quarks become efficiently thermalised. Then, only the remaining $N$'s will carry a significant boost. If these then undergo efficient $B-L$ violating hard scattering, the negative $B-L$ hidden in these states will be transferred to the visible sector, washing out the asymmetry. The inverse scattering process, which would produce boosted $N$'s (in a potentially CP violating way), is kinematically blocked because the quarks have become thermalised and lost their boost. 

\begin{figure*}[t]
\begin{center}
\includegraphics[width=230pt]{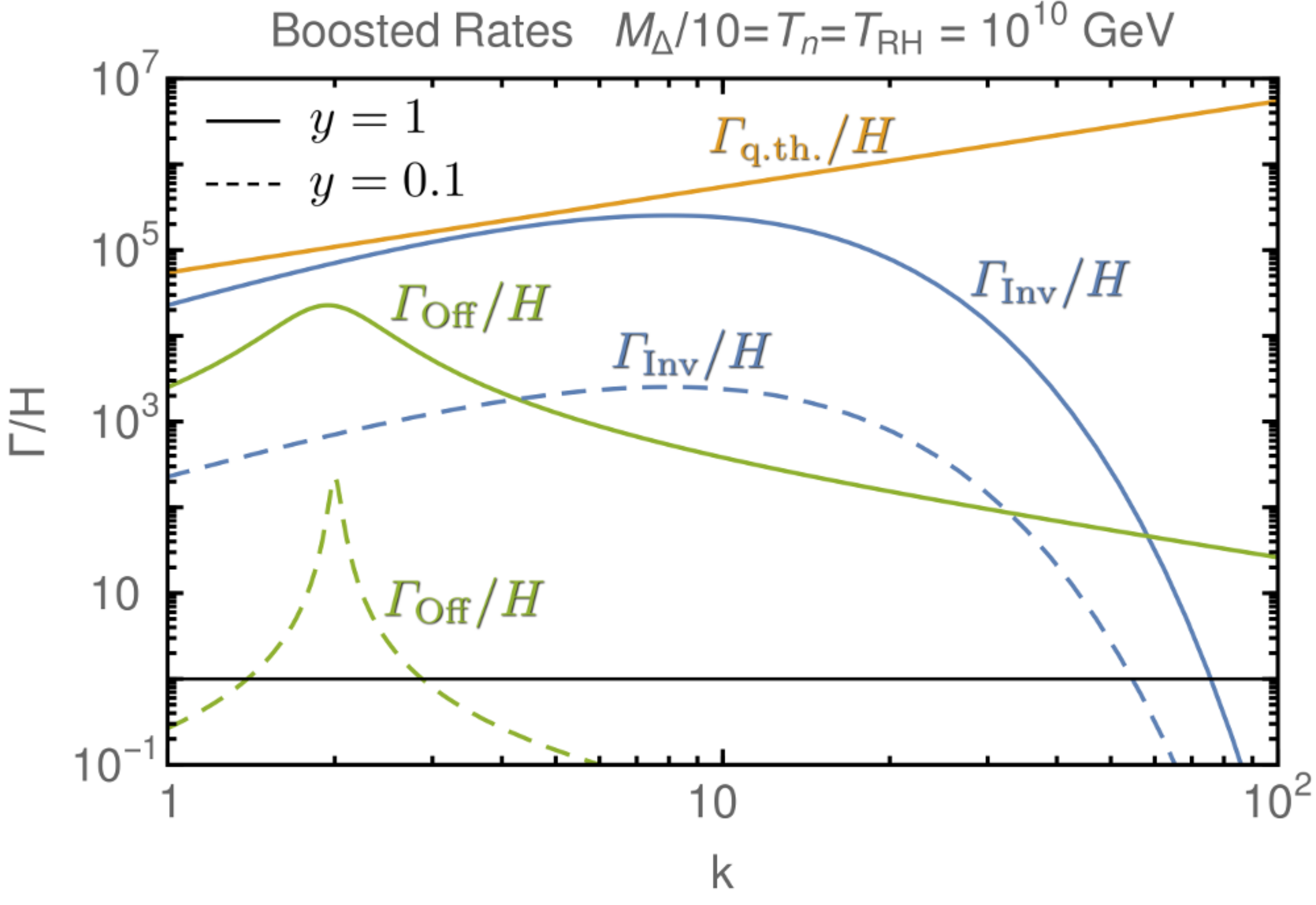} $\qquad$
\includegraphics[width=230pt]{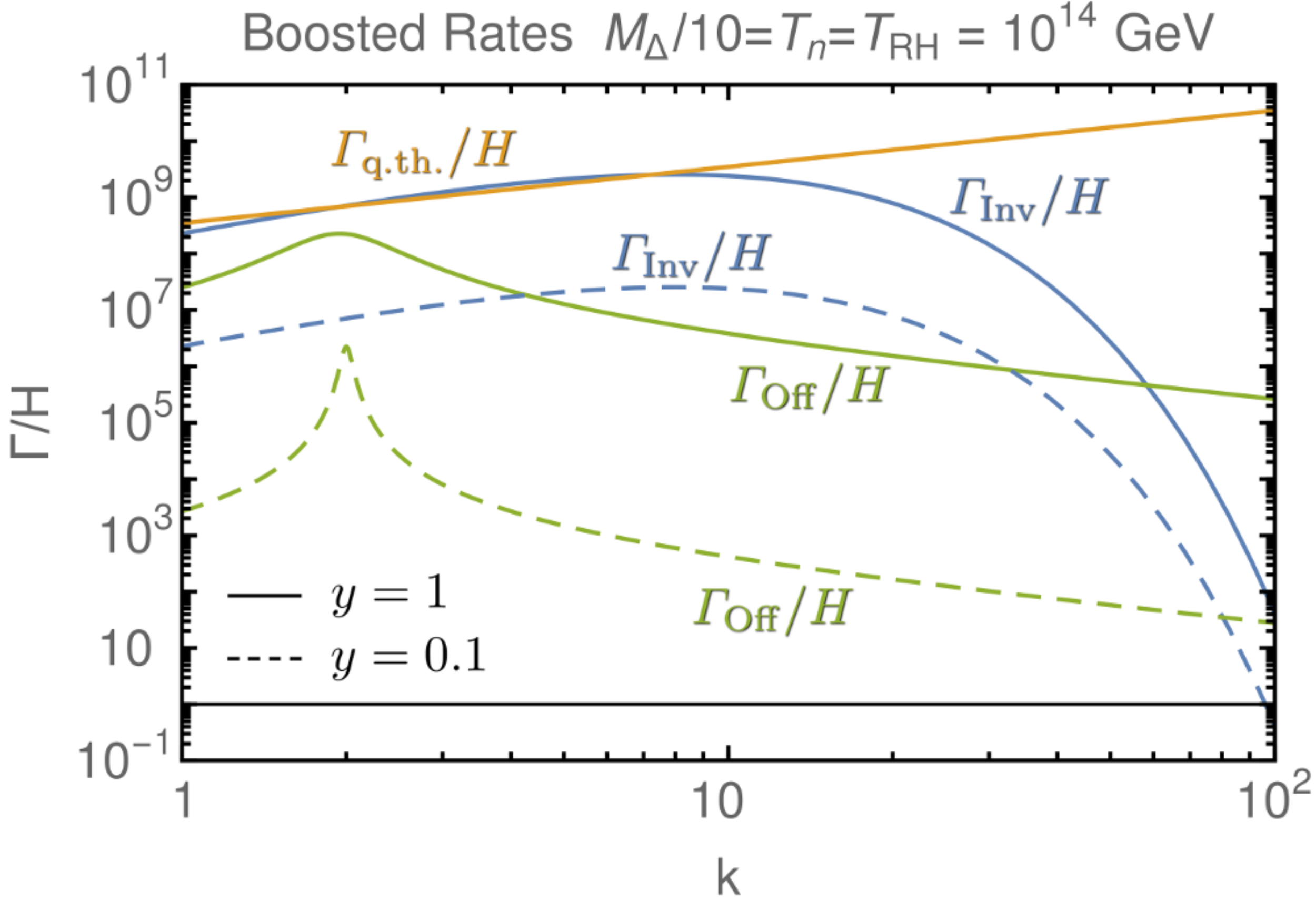}
\end{center}
\caption{Comparison of the inverse decay rate, off-shell scattering rate, and quark thermalisation as a function of $k$, where the energy of the boosted decay products is $E_{\ast} = M_{\Delta}/kT_{n}$. Unless $y \gtrsim 1$, the quark thermalisation rate is larger than the inverse decay rate. (Note the small effect of $\alpha_{s}$ running between the scales which shifts the relative size of the two rates.) The scattering rates scale as $\propto T_{n}^3$, which means they are significantly larger than $H \propto T_{\rm RH}^2$ for moderate supercooling, even at high scales.}
\label{fig:boost}
\end{figure*}

To proceed in our estimates quantifying the above picture, we start by denoting the energy of the $\Delta$ decay product as $E_{\ast} = M_{\Delta}^{2}/kT_n$, where $k \sim 2$ would correspond to the first scatterting step. First, we consider the rate of on-shell inverse decays when this boosted particle encounters the thermal bath of temperature $T_n$. The washout rate induced by such processes (e.g. $uN \to \overline{d'}\overline{d}$), is approximately 
	\begin{equation}
	\label{eq:appinv}
	\Gamma_{\rm Inv} \approx \frac{ \Gamma_{1u}\Gamma_{1d} }{ \Gamma_{\Delta} } \frac{ M_{\Delta} T_n }{ E_{\ast}^{2} } \times \mathrm{Exp} \left[ -\frac{ M_{\Delta}^{2} }{ 4E_{\ast}T_n } \right],
	\end{equation}
where a derivation is provided in App.~\ref{app:OSWO}. Secondly, at lower centre-of-mass energies, off shell processes can also play an important role. The off-shell hard scattering rate in the plasma frame is approximately
	\begin{align}
	\label{eq:Hd}
	\Gamma_{\rm Off} & = n_{d}v_{\rm rel}\sigma(dd' \to \overline{u}\overline{N}) \\
	 & \approx \frac{y^{4}n_{d}}{64\pi} \frac{s}{[(s-M_{\Delta}^{2})^{2}+M_{\Delta}^{2}\Gamma_{\Delta}^{2}]} \\
	& \approx \frac{3\zeta(3)y^{4}}{32\pi^3} \frac{E_{\ast}T_{n}^4}{[(2E_{\ast}T_{n}-M_{\Delta}^{2})^{2}+M_{\Delta}^{2}\Gamma_{\Delta}^{2}]}.
	\end{align}
Similar rates of course apply for time reversed, CP conjugate processes, and t-channel diagrams (albeit with no resonance).\footnote{The s-channel estimate is expected to break down close to the $\Delta$ peak, which corresponds to $E_{\ast} = M_{\Delta}^{2}/2T_n$ (i.e.~$k=2)$, because we have not taken into account the precise thermal distribution of the plasma particles. Nevertheless, for $k$ a little larger than two, it should provide a suitable estimate. The resonance is anyway captured by Eq.~\eqref{eq:appinv} which we shall see gives the dominant $B-L$ violating rate. For the off-shell process, we will check whether the average energy decay products interacting with the average energy thermal bath particles do or do not lead to washout, rather than taking into account precise distributions. Also note for the thermal plasma, we always write $\langle E \rangle \approx T$ to avoid some clutter, although $\langle E \rangle \approx 3T$ is more precise. But if we include the factor of three from the start, we also find $\gamma_{\Delta} \approx M_{\Delta}/3T$, so this does not spoil our overall picture.} Finally, if our initial state is a boosted quark, we should also consider the thermalisation rate of this particle via soft gluon exchange, 
	\begin{equation}
	\Gamma_{\rm q.th.} =\frac{ g_{\mathrm{QCD}\ast}\zeta(3) }{ \sqrt{2\pi^3} } \frac{ \alpha_{s}^{3/2}T_{n}^{2}}{E_{\ast}},
	\end{equation}
where the estimate follows from the same logic which leads to Eq.~\eqref{eq:kineqdelta}. To gain a clearer picture, we plot these rates normalized to Hubble for some choices of the Yukawa couplings and reheating scales in Fig.~\ref{fig:boost}. From the figure, we see the quarks will be efficiently thermalised, while any non-thermalised particles --- such as the $N$'s in the absence of additional interactions --- would undergo rapid $B-L$ violating interactions. 

To understand these results, it is instructive to consider the ratio of quark thermalisation to inverse decay rate,
	\begin{align}
	\frac{ \Gamma_{\rm q.th.} }{ \Gamma_{\rm Inv} }  & \approx \frac{ 16 \zeta(3) g_{\mathrm{QCD}\ast}  \alpha_{s}^{3/2} }{ \sqrt{8 \pi}} \frac{e^{k/4} }{ y^{2} k } \\
	& \approx \left( \frac{\alpha_s}{0.03} \right)^{3/2} \left( \frac{ g_{\mathrm{QCD}\ast} }{ 79 } \right) \times \frac{e^{k/4} }{  y^{2}k }.
	\end{align}
For $y \lesssim 1$ the quarks are therefore, on average, thermalised before creating (significant) additional $\Delta$'s. Similarly, for $k$ only a little away from two, the width in the propogator in the off-shell scattering becomes negligible compared to the first term, and the ratio of thermalisation to off-shell hard scattering rate is
	\begin{align}
 	\frac{ \Gamma_{\rm q.th.} }{ \Gamma_{\rm Off} }  & \approx \frac{ 32\pi^3 g_{\mathrm{QCD}\ast} \alpha_{s}^{3/2} }{ 3 \sqrt{2\pi^{3}} y^{4} } (2-k)^{2}  \\
	& \approx \frac{17}{y^{4}} \left( \frac{\alpha_{s}}{0.03} \right)^{3/2}\left( \frac{ g_{\mathrm{QCD}\ast} }{ 79 } \right)(2-k)^{2}
	\end{align}
So the majority of the quarks are immediately thermalised rather than undergoing hard scatterings, as we have already seen in Fig.~\ref{fig:boost}.

The $N$'s, however, have no SM gauge interactions. Let us now see what occurs if they also have no hidden sector interactions. Indeed the $N$'s will undergo $B-L$ violating interactions until the rate of their scatterings, drops below $H$. As can be deduced from Fig.~\ref{fig:boost}, this typically only occurs for large $k$. Quantifying this for the inverse decays, we find (assuming vacuum domination)	
	\begin{align}
	\label{eq:InvkH}
k & \gtrsim 4 \, \mathrm{Log} \left[ \frac{ k^{2} y^{2} T_{n}^{3} M_{\rm Pl}}{ 8\pi M_{\Delta}^{2}v_{\phi}^{2} } \sqrt{ \frac{ 3 }{ 8\pi c_{\rm vac} } } \right] \\
 	  & \approx 52 +\mathrm{Log}\left[ \frac{ y^{2} }{ \sqrt{c_{\rm vac}} } \left( \frac{k}{52} \right)^2 \left( \frac{ T_n^3 }{ 0.1^3 M_{\Delta} v_{\phi}^2 } \right) \left( \frac{10^{12} \; \mathrm{GeV} }{ M_{\Delta} } \right)  \right]. \nonumber
	\end{align}
For the off-shell scattering, on the other hand, the rate drops below $H$ only when 
	\begin{align}
	k & \gtrsim \sqrt{\frac{3}{8\pi}}\frac{ 3\zeta(3)y^{4}M_{\rm Pl}T_{n}^{3} }{32 \pi^3 \sqrt{c_{\rm vac}} M_{\Delta}^{2}v_{\phi}^2} \\
	  & \approx 46 \frac{y^{4}}{\sqrt{c_{\rm vac}}} \left( \frac{T_n^{3}}{0.1^{3}M_{\Delta}v_{\phi}^{2}} \right) \left( \frac{10^{12} \; \mathrm{GeV} }{M_{\Delta}} \right). \label{eq:OffkH}	
	\end{align}
If $y$ is somewhat suppressed, which we will see is preferred in the MG mechanism, the RHS of Eq.~\eqref{eq:OffkH} can immediately be below two and therefore the off shell-scattering can be cosmologically insignificant. The inverse decay is rapid, however, for many scattering steps as captured in Eq.~\eqref{eq:InvkH}. And therefore dangerous from the washout perspective unless $y$ is very much suppressed. 

We therefore need to assume additional dark sector interactions are present, which allow for dominant $B-L$ conserving interactions for the $N$'s in their path back to equilibrium. (These new interactions should be more rapid than the dangerous $Nu \to \Delta$ inverse decay.) Otherwise, as argued above, the boosted quarks will quickly thermalise (preserving the asymmetry), while the $N$'s undergo $B-L$ violating hard scatterings. The latter would effectively transfer the negative $B-L$ stored in the $N$'s back into the SM sector. Consequently, no asymmetry would then survive.

In Sec.~\ref{ref:Nnature} below, we shall see that $N$ may be part of an asymmetric DM sector, which means it necessarily have some additional interactions. The evaluation of the return to kinetic equilibrium in the presence of the additional asymmetric DM interactions, which allow for decays into and scattering with the other dark sector particles, is left for further work. (The precise calculation is challenging, due to the highly out-of-equilibrium nature of the $N$'s. A detailed evaluation may be possible through a monte-carlo simulation.) Finally we remark that if all the boosted decay products carry gauge interactions, as is the case in the leptogenesis option sketched in App.~\ref{sec:LGEN}, the issue of washout can also be alleviated.

\underline{Case B: $\Delta$ decays at rest}
If the $\Delta$'s decay at rest, then the decay products carry energy $E_{\ast} \approx M_{\Delta}/2$. The scattering back via the $\Delta$ resonance is further suppressed, and we are safer from washout than in case A. Let us now denote the energy of the decay products as $E_{\ast} = M_{\Delta}/k$, with $k\approx 2$ corresponding to the first scattering step. For the inverse decay, $Nu \to \Delta$, the washout rate is below $H$ for
	\begin{align}
	k & \gtrsim \frac{4 T_{n}}{M_{\Delta}} \, \mathrm{Log}\left[ \frac{y^{2}k}{8\pi} \sqrt{ \frac{3}{8\pi c_{\rm vac}} } \frac{M_{\rm Pl} T_{n}}{v_{\phi}^{2}} \right] \\
	  & \approx 2 \left( \frac{ 10 \, T_n }{ M_{\Delta} } \right) \\
	  & \; \; \times \left\{ 1 + \mathrm{Log}\left[ \frac{k}{2\sqrt{c_{\rm vac}}} \left( \frac{y}{0.1} \right)^2 \left( \frac{ 10 \, T_n }{ v_{\phi} } \right) \left( \frac{ 10^{12} \; \mathrm{GeV} }{ v_{\phi} } \right) \right] \right\}. \nonumber
	\end{align}
The off-shell hard scattering $B-L$ violating scattering for the $N$ is smaller than $H$ for 
	\begin{align}
	k & \gtrsim \sqrt{\frac{3}{8\pi}} \frac{3 \zeta(3) y^{4} T_{n}^4 M_{\rm Pl} }{ 32 \pi^3 \sqrt{c_{\rm vac}} M_{\Delta}^{3}v_{\phi}^2 } \\
	 & \approx 2 \times \frac{y^{4}}{\sqrt{c_{\rm vac}}} \left( \frac{T_n^{4}}{0.1^{4}M_{\Delta}^{2}v_{\phi}^{2}} \right) \left( \frac{10^{12} \; \mathrm{GeV} }{M_{\Delta}} \right).	
	\end{align}
Note in this case, $y \lesssim 0.1$ given the RHS in Eq.~\eqref{eq:boostcond}, so we are safer from washout. We can anyway also assume some additional interactions for the $N$, as above, to bring them safely back toward equilibrium.

\subsection{At $T_{\rm RH}$: just after bubble collision and reheating}
\label{sec:avoidingwashoutB}
\begin{figure}[t]
\begin{center}
\includegraphics[width=240pt]{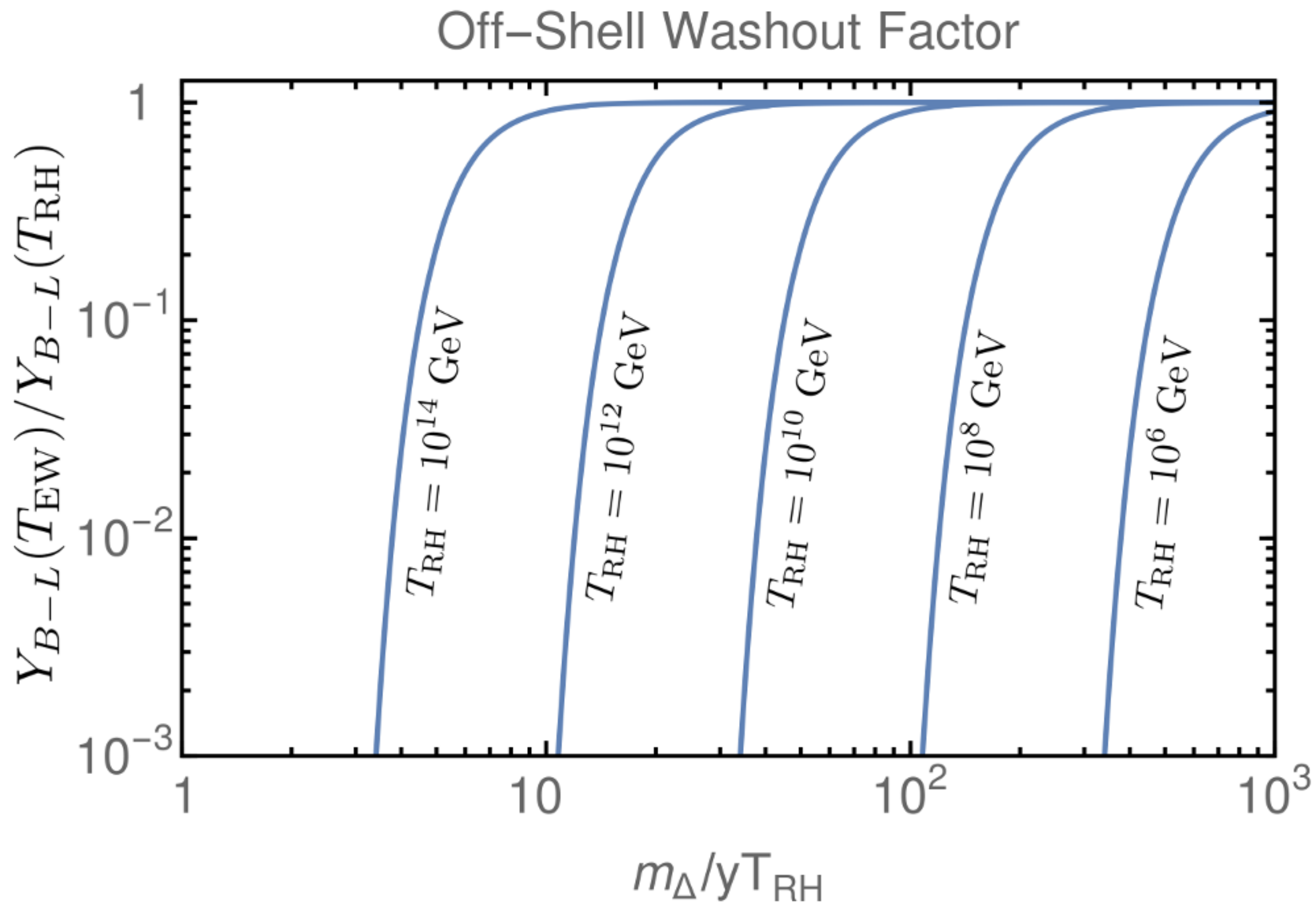}
\end{center}
\caption{Washout of the initial asymmetry, here set to $Y_{B-L}(T_{\rm RH}) = 10^{-9}$, from off-shell scatterings, as a function of the $M_{\Delta}/yT_{\rm RH}$ ratio. We assume the minimal number of quark couplings to the $\Delta$'s, see below Eq.~\eqref{eq:couplings}, and that all quarks are in chemical equilibrium. If more couplings are of the same order, then the washout rate will be enhanced, through a combinatorical factor. 
}
\label{fig:WO}
\end{figure}

Above we have dealt with washout processes before the return to equilibrium. Now we consider washout after equilibrium has been reached.
Following bubble percolation, the scalar condensate oscillates and decays, transferring its energy (negligible or otherwise) back into the plasma. The temperature of the universe is at $T_{\rm RH} \simeq \mathrm{Max}[T_n,T_{\rm infl}]$. Quark interactions violating $B-L$ can result in washout if $T_{\rm RH}$ is too high. Off shell $2 \leftrightarrow 2$ processes mediated by $\Delta$ have a rate
	\begin{equation}
	\Gamma_{\rm WO} \approx \frac{y^{4} T_{\rm RH}^{5} }{8\pi M_{\Delta}^{4} },
	\label{eq:WOguess}
	\end{equation} 
where the $8\pi$ is a mere representative guess. This rate is safely below $H$, provided that
	\begin{align}
	\frac{ M_{\Delta} }{ T_{\rm RH} }  \gtrsim  \left( \frac{ y^{4} M_{\rm Pl} }{8\pi g_{\ast}^{1/2}T_{\rm RH}} \right)^{1/4}
	 \approx 15 \times y \times \left( \frac{ 10^{12} \; \mathrm{GeV} }{  T_{\rm RH} }  \right)^{1/4}. 
	\label{eq:est}
	\end{align} 
In order to better test our estimate, we integrate out the $\Delta_{i}$ from Eq.~\eqref{eq:couplings}, and consider the induced four-fermion operators.  We solved the associated Boltzmann equations governing the washout, and calculated the suppression of the initial $Y_{B}(T_{\rm RH})$. Details can be found in App.~\ref{app:Bmann}. The results are shown in Fig.~\ref{fig:WO}, and show good agreement with the estimate in Eq.~\eqref{eq:est}. 

\begin{figure*}[t]
\begin{center}
\includegraphics[width=230pt]{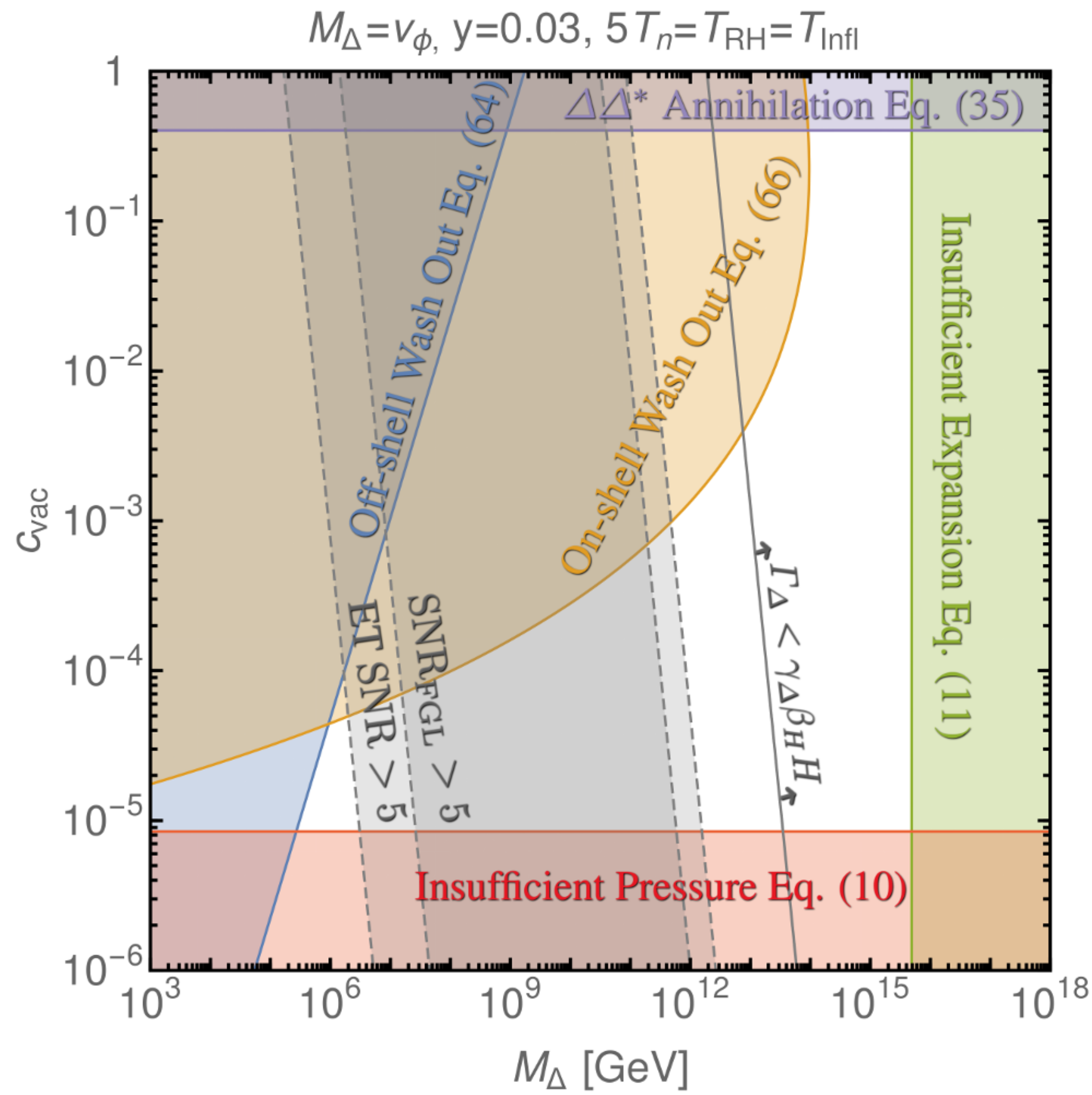}
\includegraphics[width=230pt]{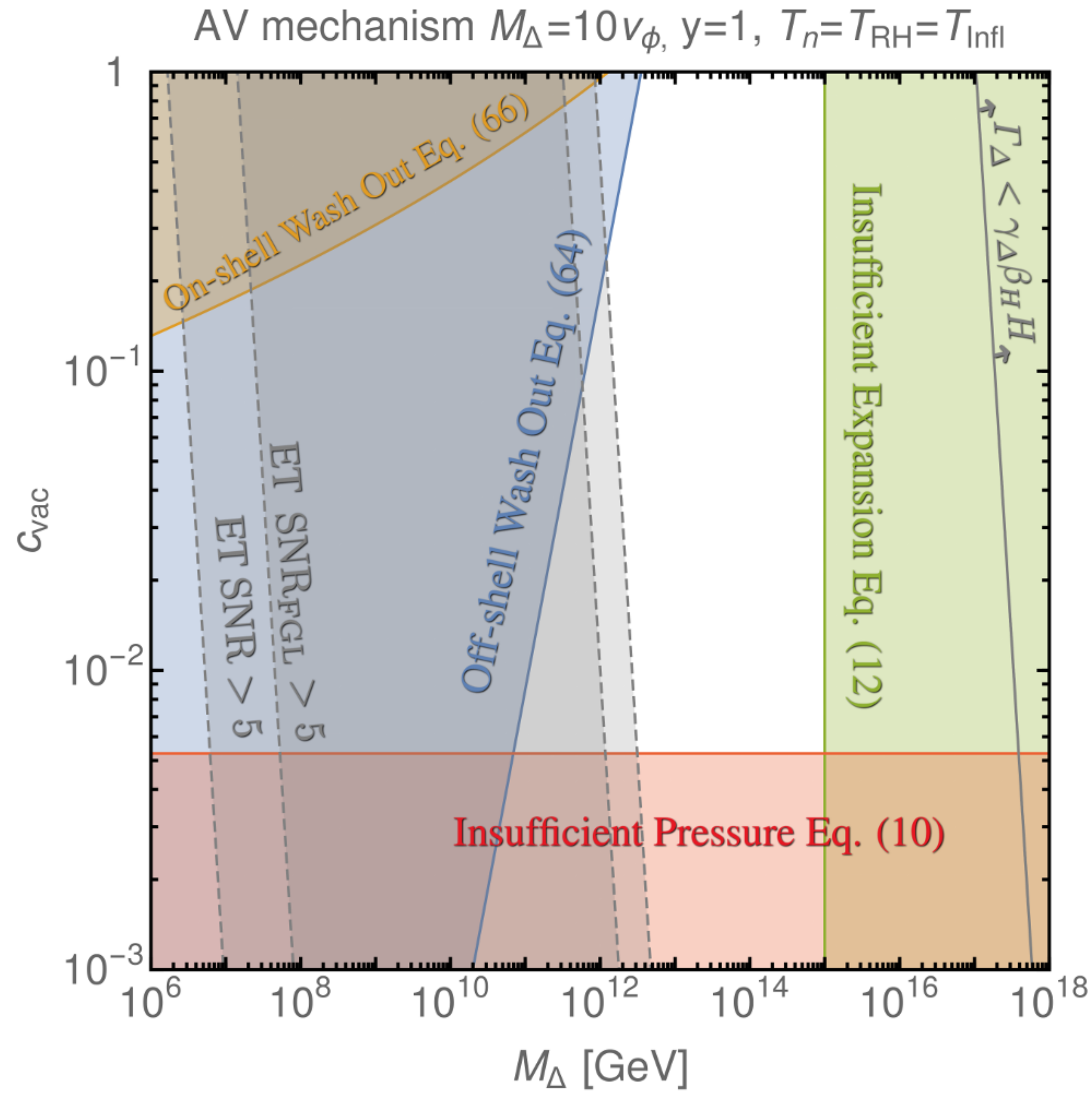}
\end{center}
\caption{Left: Constraints on the parameter space for MG-type producation.  White areas are allowed and can accommodate the baryon asymmetry. A $y$ smaller than unity helps to avoid washout. A $T_{n}$ somewhat smaller than $T_{\rm RH}$ opens up lower $c_{\rm vac}$ and $M_{\Delta}$ values of parameter space. Although we extend our plot to small $c_{\rm vac} \lesssim 10^{-3}$, in order to be as general as possible, this may be difficult to justify from a microphysical theory which does not become stuck in the metastable state (but also see main text for a possible exception). To the right of the solid gray line the $\Delta$ particles decay in the opposing bubble. Using the expected bulk properties of the phase transition, $\alpha_{\rm GW} \gg 1$, $\beta_{H} \approx 10$, we show parameter space testable by the Einstein Telescope in grey (with and without astrophysical foregrounds). See Sec.~\ref{sec:GW} for details. Right: Constraints on the parameter space for AV-type producation and for favourable assumptions $T_{n} = T_{\rm infl}$, $y =1$, and $M_{\Delta 2} \sim M_{\Delta 1} \approx 10v_{\phi}$. Here the asymmetry is suppressed by the reduced $\Delta$ yield, so we are more constrained in our choice of $y$ and $T_{n}$. White areas are allowed and can accommodate the asymmetry provided $\lambda \gtrsim 1$. Parameter space testable with the Einstein Telescope, assuming $\alpha_{GW} \approx 1$, $\beta_{H} \approx 10$, is shown in grey.}
\label{fig:sum}
\end{figure*}

Inverse decays into on-shell $\Delta$ can also lead to washout. The Boltzmann suppressed rate is given by
	\begin{equation}
	\Gamma_{\rm ID} \approx \frac{3y^{2}}{16\pi}M_{\Delta} \left( \frac{ M_{\Delta} }{ T_{\rm RH} } \right)^{3/2} \mathrm{Exp}\left[-\frac{ M_{\Delta} }{ T_{\rm RH} }\right].
	\end{equation}
This is safely below $H$, provided that
	\begin{align}
	\frac{ M_{\Delta} }{ T_{\rm RH} } & \gtrsim \mathrm{Log}\left[ \frac{ y^{2}M_{\rm Pl} }{ 8\pi T_{\rm RH } } \left( \frac{ M_{\Delta} }{ T_{\rm RH} } \right)^{\frac{5}{2}} \right] \\
	& \approx 20 + 2 \, \mathrm{Log}[y] + \frac{5}{2}\mathrm{Log}\left[\frac{ M_{\Delta} }{ 20  T_{\rm RH}}\right] - \mathrm{Log}\left[ \frac{  T_{\rm RH} }{ 10^{12} \; \mathrm{GeV} } \right]. \nonumber
	\end{align}

\subsection{Short summary of the results}

Having found the dependence of the yield on the mass spectrum, properties of the phase transition, CP violating couplings, and discussed the conditions needed to avoid washout, we now summarize our findings in Fig.~\ref{fig:sum}. As can be seen from the figure, washout can be avoided and the baryon asymmetry explained, both for the MG and AV-type mechanisms. For more realistic values, $c_{\rm vac} \gtrsim 10^{-3}$, which are typical for close-to-conformal potentials and can accommodate sufficient bubble nucleation, we find $10^{12} \; \mathrm{GeV} \lesssim M_{\Delta} \lesssim 10^{15} \; \mathrm{GeV}$ for both MG and AV-type mechanisms.

In order to cast as wide a net as possible, we also extend our results to smaller $c_{\rm vac}$ --- which is consistently possible for the MG mechanism --- even with some suppression of $Y_{B}$ through $y \sim 10^{-2}$ and $T_{n} \sim T_{\rm RH}/5$. Sufficient bubble nucleation in order for the phase transition to complete may still be possible in this regime. For example, if $\phi$ is linked through a Yukawa to some hidden strong sector fermions undergoing condenstation, then bubble nucleation can be massively enhanced below the confinement temperature~\cite{Witten:1980ez,Spolyar:2011nc,Iso:2017uuu,vonHarling:2017yew}. Such model building is left for future work.

\section{Gravitational wave signal}
\label{sec:GW}

For the gravitational wave signal we use the latest --- state of the art --- numerical results for thick walled bubbles calculated by Cutting et al.~\cite{Cutting:2020nla}. (For related studies see~\cite{Jinno:2017fby,Konstandin:2017sat,Cutting:2018tjt,Lewicki:2020jiv}.) The spectrum depends on the bulk parameters of the phase transition. Namely: 
	\begin{itemize}
	\item The proportion of the energy going into the scalar contribution, for us $\kappa_{\phi} \simeq 1$.
	\item The wall velocity which for our runaway transitions is $v_{w}\simeq 1$.
	\item The Hubble scale at the transition, $H_{\ast}$, which is easily found given $T_{\rm n}$ and $T_{\rm RH}$.
	\item The bubble size at collision, 
	\begin{equation}
	R_{\rm col}H_{\ast} \approx \frac{(8\pi)^{1/3}v_{w}}{\beta_H}.
	\end{equation}
This is in turn set by the parameter
	\begin{equation}
	\beta_{H} \equiv \frac{\beta}{H_{\ast}} = -T\frac{dS}{dT},
	\end{equation}
evaluated at $T_{\rm n}$, which was introduced earlier, but of which we remind the reader here.
	\item And the energy released into bulk motion during the transition,
	\begin{align}
	\alpha_{\rm GW} = \frac{1}{\rho_{\rm rad}}\Delta\left( V_{\rm eff} - \frac{T}{4} \frac{ \partial V_{\rm eff} }{ \partial T }\right),
	\end{align} 
where the difference is calculated between the two minima of the effective potential at nucleation. And which is normalised by convention to the radiation density in the plasma $\rho_{\rm rad}$.
	\end{itemize}
The latest results give a gravitational wave spectrum as measured today, for initially thick walled, runaway bubbles~\cite{Cutting:2020nla}
	\begin{align}
	h^2\Omega_{\rm GW}(f) & \equiv h^2\frac{d\Omega_{\rm GW}}{d \mathrm{log}(f)} \nonumber \\
	& = 4.4 \times 10^{-7} \left( \frac{H_{\ast}}{\beta} \right)^{2} \left( \frac{\alpha_{\rm GW}}{1+\alpha_{\rm GW}} \right)^{2} \\
	& \qquad \times  \left( \frac{g_{\ast}(T_{\rm RH})}{100} \right)^{-1/3}   S_{\phi}(f). \nonumber
	\end{align}
In converting the expression in~\cite{Cutting:2020nla} to be in terms of $\alpha_{\rm GW}$, we have assumed the change in radiation density between the two phases is negligible compared to the vacuum energy (well justified here). The spectral shape is described by
	\begin{equation}
	S_{\phi}(f) = \frac{(a+b)\tilde{f}^bf^a}{b\tilde{f}^{(a+b)}+af^{(a+b)}},
	\end{equation}
where we use the central values for the bubbles with the thickest walls from~\cite{Cutting:2020nla}, namely $a=0.742$ and $b=2.16$.
Here the peak frequency redshifted to today is
	\begin{equation}
	\tilde{f} = 330 \; \mathrm{Hz} \, \left( \frac{g_{\ast}(T_{\rm RH})}{100} \right)^{1/6} \left( \frac{\beta}{H_{\ast}} \right) \left( \frac{ T_{\rm RH} }{ 10^{10} \; \mathrm{GeV} } \right).
	\end{equation}	
Superhorizon GW modes entering the horizon during the radiation dominated phase following the phase transition are known to result in a spectrum $\Omega_{\rm GW} \propto f^{3}$ in the IR~\cite{Durrer:2003ja,Caprini:2009fx,Barenboim:2016mjm,Cai:2019cdl,Hook:2020phx}. This is not captured by the above simulations. We therefore impose a cut in the spectrum at the redshifted frequency today of
	\begin{align}
	f_{\ast}(t_0) & = \left(\frac{ a_{\ast}}{a_0} \right) \times \frac{ H_{\ast}}{ 2\pi } \\
	 & =  260 \; \mathrm{Hz} \left( \frac{g_{\ast}(T_{\rm RH})}{100} \right)^{1/6} \left( \frac{ T_{\rm RH} }{ 10^{10} \; \mathrm{GeV} } \right), \nonumber
	\end{align}
below which we enforce the correct scaling (here $a_{\ast}/a_0$ is the ratio of scale factors between the phase transition and today). 

Of the current or upcoming gravitational wave interferometers, our model can best be probed using the Einstein Telescope (ET), due to its sensitivity at high frequencies. The sensitivity of the ET to a GW signal is given by
	\begin{equation}
	h^2\Omega_{\rm sens}(f) = \frac{2\pi^{2}f^{3}}{3H_{100}^2}S_{n}(f),
	\end{equation}
where $H_{100} = 100 \, \mathrm{km}/\mathrm{s}/\mathrm{Mpc}$, and $S_n$ is the noise spectral density~\cite{Hild:2010id}. For a long lasting stochastic source, the signal-to-noise ratio is~\cite{Thrane:2013oya}
	\begin{equation}
	\mathrm{SNR} = \sqrt{t_{\rm obs}\int df \left(\frac{\Omega_{\rm GW}(f)}{\Omega_{\rm sens}(f)}\right)^2},
	\end{equation}
where $t_{\rm obs}=10$ years is our fiducial choice for the ET observation time. Using this expression, so called power-law-integrated (PLI) sensitivity curves are calculated~\cite{Thrane:2013oya}, and our calculation for the ET PLI sensitivity curve is shown in our figures. There is an astrophysical stochastic GW foreground coming from binaries with black hole and neutron star constituents in the ET frequency range. We take the estimate for the foreground, $\Omega_{\rm FG}(f)$, from the LIGO/VIRGO/KAGRA collaboration~\cite{LIGOScientific:2021iex} (specifically the median value given therein), which uses the latest observational data. Whether a given cosmological signal can be disentangled from a partially uncertain astrophysical foreground is an active area of research~\cite{Cutler:2005qq,Adams:2010vc,Adams:2013qma,Sachdev:2020bkk,Flauger:2020qyi,Boileau:2020rpg,Martinovic:2020hru,Romero:2021kby}. In order to be conservative, we also define a simple foreground limited signal-to-noise ratio,  
	\begin{equation}
	\mathrm{SNR}_{\mathrm{FGL}} = \sqrt{t_{\rm obs}\int df \left(\frac{\mathrm{Max}[\Omega_{\rm GW}(f)-\Omega_{\rm FG}(f),0]}{\Omega_{\rm sens}(f)}\right)^2}.
	\end{equation}
Using $\mathrm{SNR}_{\mathrm{FGL}}$ ensures we do not claim a signal is detectable if it buried under $\Omega_{\rm FG}(f)$.\footnote{A similar analysis for GW signals in dilaton mediated DM models, also including additional interferometers, has been performed in~\cite{Baldes:2021aph}.} In our summary plots, Fig.~\ref{fig:sum}, we show the regions of parameter space ET is sensitive to the GW signal expected in our scenario. The scenario could be tested further if GW detection experiments were extended to higher frequencies~\cite{Aggarwal:2020olq}. Astrophysically, GWs at frequencies above the ET range are not of interest, but the SM thermal plasma itself provides a cosmological source with amplitude dependent on the reheating temperature following (slow roll) inflation, and with peak frequency around $10^{11}$~Hz~\cite{Ghiglieri:2015nfa,Ghiglieri:2020mhm,Ringwald:2020ist}. Apart from the high scale PTs, as studied here, other beyond the SM scenarios can also lead to significant signals at high frequencies. These include light primordial black holes decaying prior BBN~\cite{Anantua:2008am,Fujita:2014hha,Dong:2015yjs,Inomata:2020lmk}, models of axion inflation~\cite{Barnaby:2010vf,Adshead:2019lbr}, and models producing metastable cosmic strings~\cite{Buchmuller:2019gfy}. Currently the strongest constraint at such frequencies comes from the energy density during BBN, which limits the total $\Omega_{\rm GW} \lesssim 1.2 \times 10^{-6}$ above frequencies of $10^{-15}$ Hz~\cite{Pagano_2016}. Constraints from graviton to photon conversion are orders of magnitude weaker~\cite{Domcke:2020yzq}.

\begin{figure*}[t]
\begin{center}
\includegraphics[width=240pt]{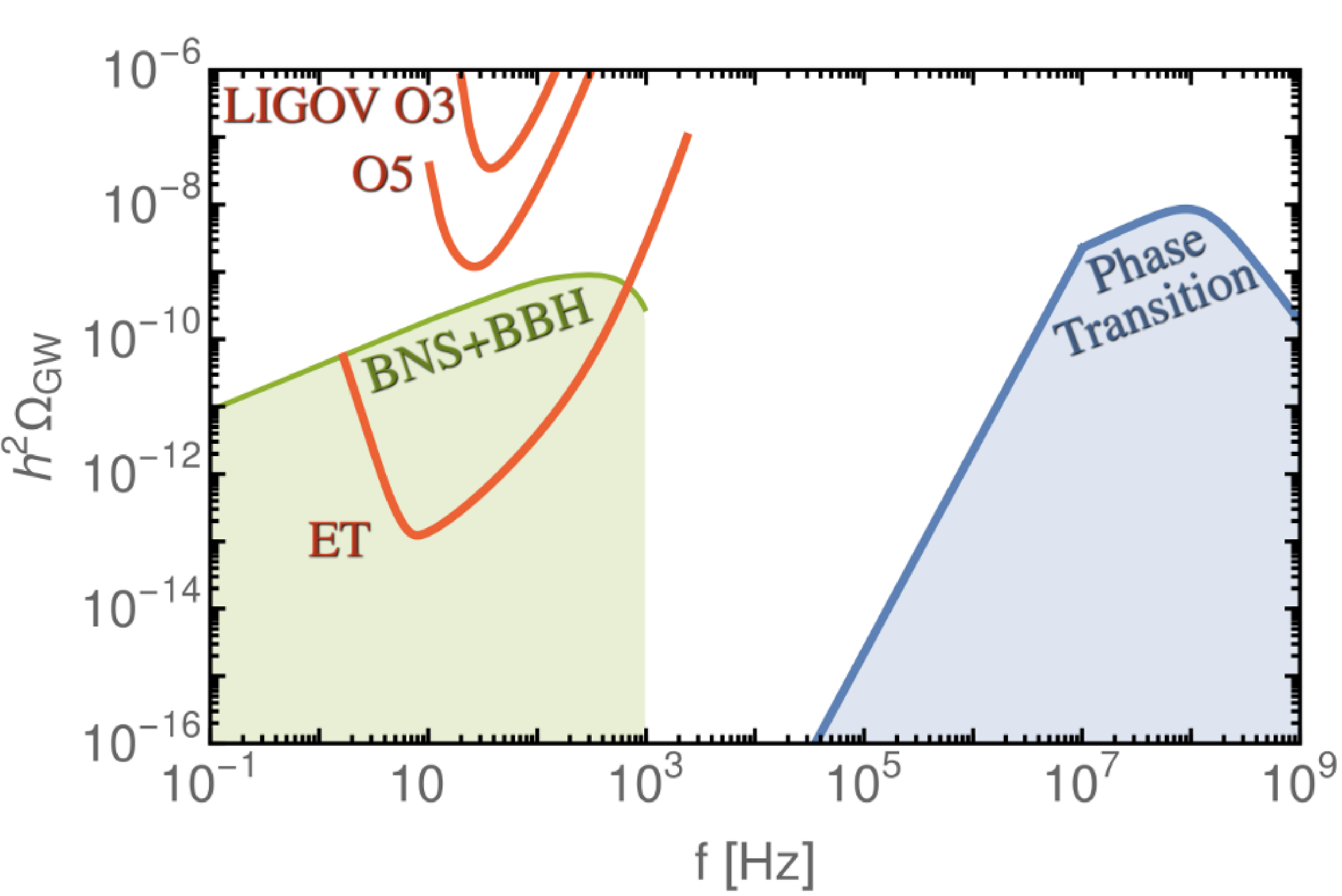}
\includegraphics[width=240pt]{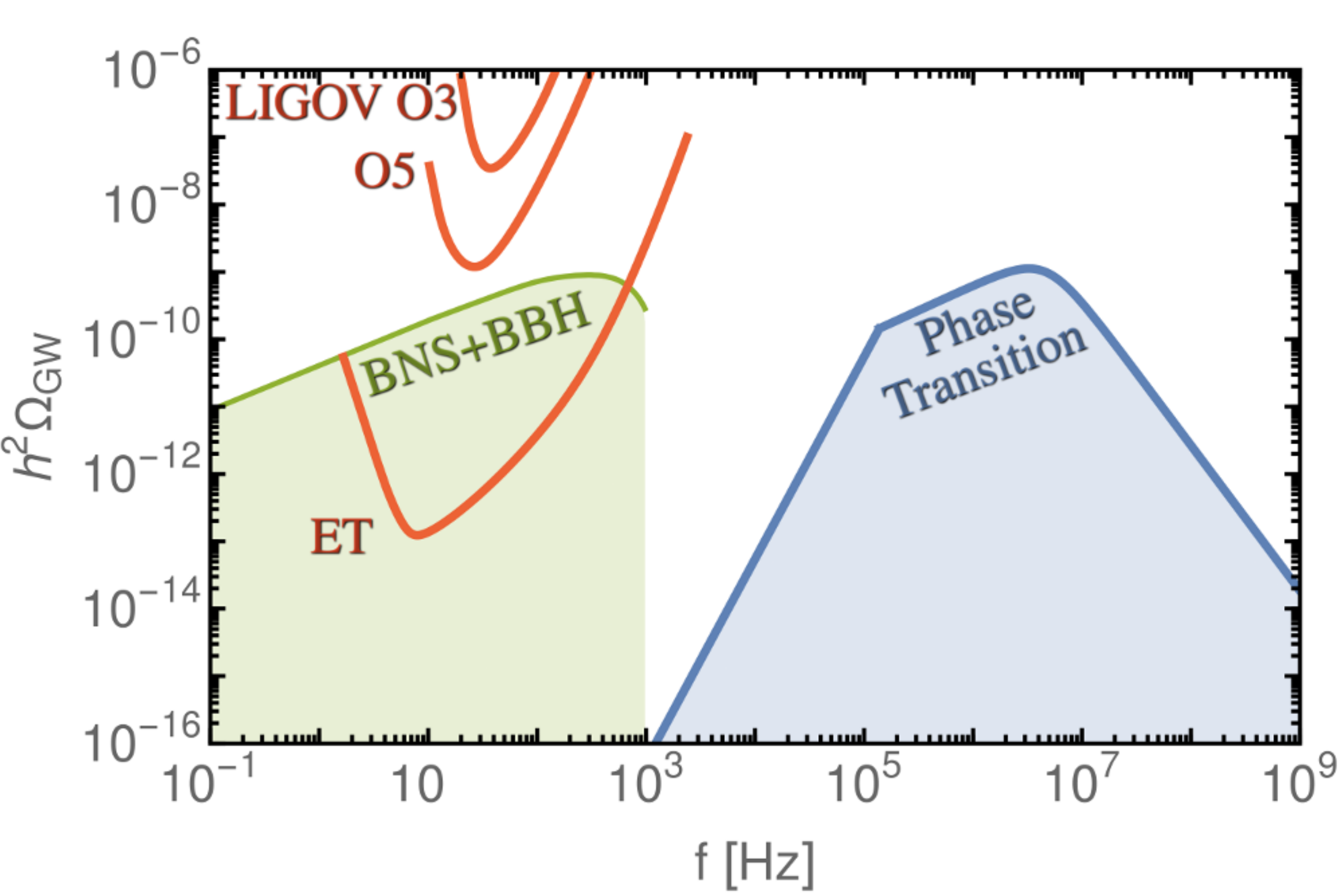}
\end{center}
\caption{Left: Example GW spectrum for the classically scale invariant potential.  The parameters chosen are $\lambda = 2.1$, $v_{\phi} = 3 \times 10^{15}$ GeV, which gives $c_{\rm vac} = 0.01$. The bulk phase transition properties are $T_{n}= 6.9 \times 10^{13}$ GeV, $T_{\rm RH}= 3.8 \times 10^{14}$ GeV, $\alpha_{\rm GW} = 930$, $\beta_H = 7.0$. Also shown are the power-law-integrated sensitivity curves for LIGO-VIRGO O3,  LIGO-VIRGO Design A+~sensitivity~\cite{LIGOScientific:2021iex}, and the ET, all with $\mathrm{SNR}=10$. The astrophyiscal foreground from binary black holes and neutron stars is 
artificially cut, due to the limited plot range in~\cite{LIGOScientific:2021iex}, but it continues to fall rapidly at higher frequencies. Right: Example GW spectrum for the potential with tree level mass terms. The parameters chosen are:  $\lambda_{\phi} = 10^{-8}$, $\lambda=2.2$, $v_{\phi} =10^{14}$ GeV, $\mu_\phi=5.3\times 10^{10}$ GeV, renormalisation scale $\mu=6.3\times 10^{13}$ GeV, the vacuum energy parameter is $c_{\rm vac} = 5.8 \times 10^{-3}$. The bulk phase transition properties are: $T_{n} = 5.2 \times 10^{12}$ GeV, $T_{\rm RH} = 1.1 \times 10^{13}$ GeV, $\alpha_{\rm GW} = 21$, and $\beta_{H} = 19$. Two field potentials, in which we can move further from close-to-conformality (tiny $\lambda_{\phi}$), are left for future work.}
\label{fig:GW}
\end{figure*}

\section{Example potentials}
\subsection{MG mechanism --- close-to-conformal potential}
For this scenario we require $M_{\Delta}\approx 0$ in the unbroken phase and a relatively strong phase transition. A simple and relatively minimal realisation is to consider a scale invariant potential at the classical level,
	\begin{equation}
	V_0(\phi,\Delta)=\frac{\lambda_\phi}{4}\phi^4+\frac{\lambda}{2}\phi^2|\Delta|^2+\lambda_\Delta|\Delta|^4.
	\end{equation}
The scale invariance is broken by the running of the couplings. We do not consider a full analysis here, e.g.~by studying the running between the Planck and IR scales for all couplings. Nor do we demand stability and the appropriate tuning for the EW scale. We simply provide a simple toy example showing the interplay between $c_{\rm vac}$, $T_{n}$, $\lambda$ and $M_{\Delta}$. Note we require two copies of $\Delta$ for CP violation, which in the classically scale invariant theory, should both gain a mass during the PT. We capture the approximate effect by include a doubling of the degrees of freedoms, captured by the factor $N_{\Delta} =2$, entering the beta and thermal functions below. Then, the beta function for the $\phi$ self coupling is given by
	\begin{equation}
	\beta_{\lambda_\phi}=\frac{1}{16\pi^2}\left(3N_{\Delta}\lambda^2+18\lambda_\phi^2\right).
	\end{equation}
Thus $\lambda_{\phi}$ runs to negative values in the IR, which triggers symmetry breaking radiatively, as first discussed by Coleman-Weinberg~\cite{PhysRevD.7.1888, Gildener:1976ih}. The coupling $\lambda_\phi$ can now be traded for the VEV $v_{\phi}$, i.e.~dimensional transmuted. The potential in the $\phi$ field direction is given by
	\begin{equation}
	V_1(\phi)=\frac{\beta_{\lambda_\phi}}{4}\phi^4\left(\log\frac{\phi}{v_\phi}-\frac{1}{4}\right).
	\end{equation}
Note our parameter $c_{\rm vac}$ is now determined by the beta function. Near the symmetry breaking scale, $\lambda_\phi \simeq 0$, so the beta function is set by the portal coupling $\lambda$. To the above potential, we add the usual one-loop thermal contribution, arising from the field and temperature dependent mass
	\begin{equation}
	M_{\Delta}^{2}(\phi,T) = \frac{\lambda}{2} \phi^{2} + \Pi_\Delta(T),
	\end{equation} 
where the thermal mass arises from the $\Delta$ coupling to the gauge bosons and to $\phi$,
	\begin{equation}
	\Pi_\Delta(T)=T^2\left(\frac{1}{3}g_s^2+\frac{1}{9}g_Y^2 + \frac{1}{24}\lambda\right).
	\end{equation}
(We ignore the subdominant Yukawa contribution in the MG option and the effect of the self-quartic $\lambda_{\Delta}$.) The changing mass results in a changing pressure of the dilute weakly interacting plasma, which leads to the aforementioned thermal contribution to the effective potential
	\begin{align}
	\label{eq:ThermalPotential}
	V_T(\phi,T) & =  \frac{g_{\Delta}N_{\Delta}T^{4}}{2\pi^2} \\
		    & \times \int_{0}^{\infty} dy y^{2} \log \left(1-\mathrm{Exp}\left[ -\sqrt{1+M_{\Delta}^{2}/T^{2}} \right] \right). \nonumber
	\end{align}
By writing the above, we have implicitly implemented the Parwani method of resumming the daisy diagrams~\cite{Parwani:1991gq}, as opposed to the Arnold-Espinosa method~\cite{Arnold:1992rz}. The subdominant thermal contribution of the $\phi$ is ignored here. The effective potential we consider is then given by
	\begin{equation}
	V_{\rm eff}(\phi,T) = V_1(\phi) + V_T(\phi,T).
	\end{equation}
At zero temperature, the $\phi$ has no effective mass term at the origin of the potential. A positive thermal mass squared for the $\phi$, $\propto \lambda^{2}T^{2}$, is generated by the thermal contribution of the $\Delta$. The result is a barrier in $V_{\rm eff}$ and a strong first order phase transition. The amount of cooling below the critical temperature, $T_{c}$, before the phase transition takes place is determined by the interplay of $\lambda$ in setting $\beta_{\lambda_\phi}$ and in generating the thermal mass: a smaller $\lambda$ results in a smaller $c_{\rm vac}$ but also a smaller thermal mass. In the end, the zero temperature effect wins out, decreasing $\lambda$ ends up increasing the strength of the phase transition. 

The bubble nucleation rate per unit volume is given by
	\begin{equation}
	\Gamma_{\rm Bub.} \sim T^{4}e^{-S}.
	\end{equation}
For the amount of cooling below $T_{c}$ relevant for us, $S=S_{3}/T$, where
	\begin{equation}
	S_{3} = 4\pi \int_{0}^{\infty} dr r^{2} \left\{ \frac{1}{2}  \left(\frac{ d\phi }{dr}\right)^{2} + V_{\rm eff}(\phi) \right\}
	\end{equation}
is the O(3) symmetric Euclidean action. To find the initial bubble profile, we solve the resulting equation of motion,
	\begin{equation}
	\frac{ d^{2}\phi }{dr^2} + \frac{2}{r}\frac{ d\phi }{dr} - \frac{\partial  V_{\rm eff}}{\partial \phi} =0, \quad \frac{ d\phi(0) }{dr}= 0, \quad \phi(\infty) = 0,
	\end{equation}
numerically using a shooting algorithm. The phase transition takes place when the bubble nucleation rate catches up to Hubble, $\Gamma_{\rm Bub.} = H^{4}$, which we have taken to define $T_{n}$. 

We next compute $T_n$ for various choices of $\lambda$ and scales $v_{\phi}$. To obtain a large enough $Y_{B}$ the nucleation temperature must not fall too far below $T_{\rm RH}$. This limits our choice of the coupling to $\lambda \gtrsim 1$, which also constrains $c_{\rm vac} \gtrsim 10^{-3}$. Combining these findings with the various other constraints in Fig.~\ref{fig:sum}, limits our viable region to $M_{\Delta} \approx v_{\phi} \gtrsim 10^{13}$ GeV with this minimal realisation of the potential. This puts us out of range of the ET. (Albeit there is a large signal present which could in principle be searched for in the far future.)  An example GW spectrum for the classically scale invariant potential is shown in Fig.~\ref{fig:GW} left. The high temperatures required for our example are consistent with the observed value of the the amplitude of scalar perturbations, $A_{s}$, and current limits on the tensor-to-scalar ratio, $r$~\cite{Planck:2018jri,BICEP:2021xfz}. (The combination limits the energy density at inflation via the relation $8\pi\rho_{\inf} = 3H_{\inf}^{2}M_{\rm Pl}^{2} \approx 3\pi rA_{s}M_{\rm Pl}^{4}/16$.) Note, even with the maximally allowed reheat temperature following inflation, the GWs sourced from the SM plasma would be well below our PT signal at its peak~\cite{Ghiglieri:2015nfa,Ghiglieri:2020mhm,Ringwald:2020ist}. Note also that the low $M_{\Delta}$ and $c_{\rm vac}$ region of Fig.~\ref{fig:sum} requires a more complicated potential for viable phase transitions.

\subsection{MG mechanism --- potential with tree level mass terms}

Next, we move away from the classically scale invariant case, and allow for tree level mass terms. The tree level potential is given by
	\begin{equation}
	V_0(\phi)=-\frac{\mu_\phi^2}{2}\phi^2+\frac{\lambda_{\phi}}{4}\phi^4+\mu_\Delta^2|\Delta|^2+\lambda_\Delta|\Delta|^4+\frac{\lambda}{2}\phi^2|\Delta|^2,
	\end{equation}  
where it will be assumed that $\mu_\Delta\simeq0$ for simplicity (and as required for baryogenesis). In principle, there can also be a tree level barrier via a cubic term for $\phi$, but we do not pursue this possibility here. We now take $N_{\Delta} =1$, as the heavier $\Delta$ can already have a non-negligible mass in the unbroken phase, unlike in the scale invariant example. The one-loop zero temperature corrections are
	\begin{align}
	V_1(\phi)=&\frac{1}{64\pi^2}\left\{6M_\Delta^4(\phi)\left(\log \left[\frac{M_\Delta^2(\phi)}{\mu^2}\right]-\frac{3}{2}\right)\right.\\
	&\left.+m_\phi^4(\phi)\left(\log \left[\frac{m_\phi^2(\phi)}{\mu^2}\right]-\frac{3}{2}\right)\right\},
	\end{align}
where $\mu$ is the renormalisation scale appearing in the $\overline{\text{MS}}$ renormalisation scheme~\cite{Quiros:1999jp}. For the thermal corrections, we again include the $\Delta$, and now also the $\phi$, where the $\phi$ mass is given by
	\begin{equation}
		m_\phi^2(\phi)=3\lambda_{\phi}\phi^2-\mu_\phi^2,
	\end{equation}
with thermal mass squared
	\begin{equation}
		\Pi_\phi(T)=\left(\frac{\lambda_\phi}{4}+\lambda\right)T^2.
	\end{equation}
We again use the Parwani method of daisy resummation~\cite{Parwani:1991gq}. The thermal correction is defined as in Eq.~\eqref{eq:ThermalPotential}, but with an extra contribution coming from the $\phi$, with $g_\Delta\rightarrow g_\phi=1$ and $M_\Delta\rightarrow m_\phi$. Summing the contributions, the effective potential is given by
	\begin{equation}
	V_{\rm eff}(\phi,T) = V_0(\phi)+V_1(\phi)+V_T(\phi,T).
	\end{equation}
We again search for suitable parameter points, which return the required $c_{\rm vac}$, $T_{\rm RH}$ and $T_{n}$, following a calculation of the bounce action as above. We have again identified suitable parameter points at high scales, $T_{\rm RH} \gtrsim 10^{13}$ GeV, which enable successful baryogenesis. One such example GW spectrum is shown in Fig.~\ref{fig:GW} right. Detailed exploration of such potentials, which could in principle also include multiple scalar field directions (in which strong phase transitions are possible also with somewhat larger quartics, as soon as tree level barriers are introduced), is left for future work. In particular, it is of interest whether the $c_{\rm vac} \lesssim 10^{-3}$ region in Fig.~\ref{fig:sum} can be populated when considering concrete realisations for the scalar sector. Detailed studies of the wall elasticity would also be interesting to pursue.

\subsection{AV mechanism --- comment on potentials}

As $M_{\Delta} \neq 0 $ in the unbroken phase for AV-type production, a classically scale invariant potential is obviously not a possibility. The potential with tree level mass terms, however, can easily be adjusted to match the requirements of the AV mechanism. One simply renames $\Delta \to X$ above, where $X$ is some new scalar field with no $B-L$ violating interactions. The $X$ now provides the same thermal barrier as before, up to some modifications of its thermal mass term, depending on its underlying gauge interactions. Then the true $\Delta$ is heavier than $v_{\phi}$ and can be integrated out. The example shown in Fig.~\ref{fig:GW} right, also covers this example, up to some small modifications. Of course, an extended analysis of the possibilities, as outlined above, can also be performed for the AV-type production mechanism.

\section{Detailed nature of $N$}
\label{ref:Nnature}
So far we have left the detailed low energy nature of $N$ unspecified. Its role so far has been to carry (hide) the negative $(B-L)$ asymmetry of the universe. We now discuss the constraints on its nature in order for it not to transfer the negative $(B-L)$ back into the visible sector and therefore spoil our baryogenesis mechanism.
\subsection{Super light $N$}
The simplest realisation is for $N_R$ to be (close-to)-massless so that it functions as dark radiation until very late times. Indeed, in this case, one may couple it to the SM Higgs and lepton doublets, $y_{\nu}\overline{l_{L}}HN_{R}$, and explain the neutrino masses with $y_{\nu}\sim 10^{-12}$. Such a coupling realises the scenario of Dirac leptogenesis~\cite{Dick:1999je}. The asymmetry is preserved because the $B-L$ number stored in the $N_{R}$ is not transferred to the active neutrinos while the sphalerons are active. (We still require some addition interaction of the $N_{R}$ for it to return to equilibrium without inducing washout following the phase transition.)

\begin{figure}[t]
\begin{center}
\includegraphics[width=240pt]{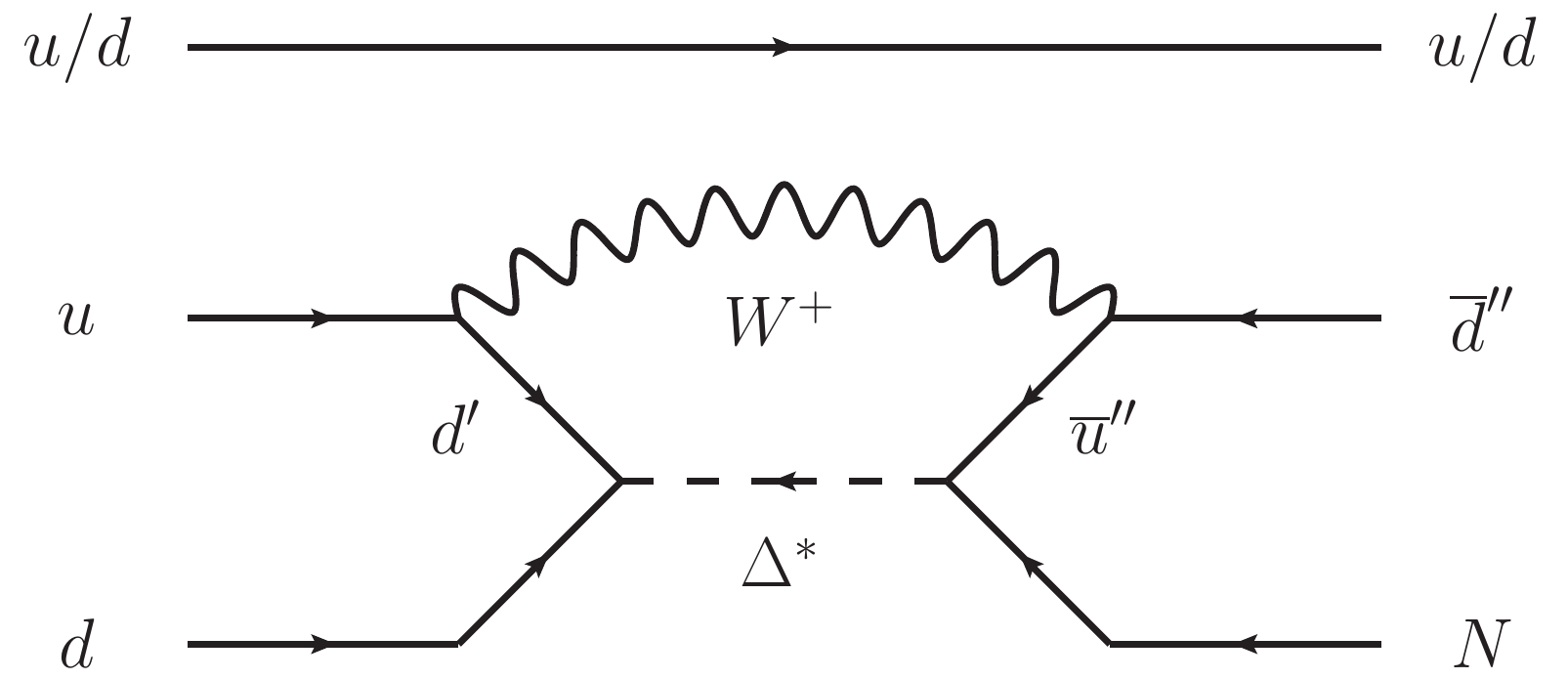}
\end{center}
\caption{Example of a Feynman diagram showing nucleon decay induced by $\Delta$ in the case when $N$ is close to massless. Such couplings are most constrained by the limits on $n \to \pi^0 \nu$ and $p \to K^+ \nu$. Some other diagrams also contribute at a similar rate.
}
\label{fig:Ndec}
\end{figure}

As $N$ is very light, however, nucleon decay is suppressed only by the off-diagonal nature of the $y_{di}$ couplings together with the large $M_{\Delta}$ scale. A representative Feynman diagram is shown in Fig.~\ref{fig:Ndec}. Taking the couplings to all generations to be of the same order, we estimate the decay rate 
	\begin{equation}
	\Gamma(N \to \pi/K+\nu) \sim \frac{y^{4}g_2^{4}|V_{ud'}^{\ast}V_{u''d''}|^{2}m_{d'}^{2}m_{u''}^{2}M_{N}^{5}}{\mathrm{Max}[M_{W}^{4},M_{u''}^{4}] \times M_{\Delta}^{4}},
	\end{equation}
where $g_2$ is the SU(2) gauge coupling and the quark masses in the numerator enter due to the chirality flips required between the $W$ boson vertex and the $\Delta$ which interacts with the right handed quark sector. From the experimental limits on $n \to \pi^{0} \nu$ of $\tau > 1.1\times 10^{33}$ years~\cite{Abe:2013lua} and $p \to K^{+} \nu$ of $\tau > 5.9 \times 10^{33}$ years~\cite{Abe:2014mwa}, we find the bound on the mass scale then reads $M_{\Delta} \gtrsim y \times (10^{12}-10^{13}) \; \mathrm{GeV}$, depending on which couplings $y$ enter.  Comparing to Fig.~\ref{fig:sum}, some of the lower range of the allowed parameter space is thus ruled out for democratic couplings, if the $N$ is indeed light.

If the coupling to the first generation is suppressed, an additional $W$ boson must be included in the decay diagram. The limit is weakened only a little, however, and then reads $M_{\Delta} \gtrsim y \times 10^{12}$ GeV.

\subsection{Massive $N$}
Now consider, instead, a relatively massive $N$. The couplings of Eq.~\eqref{eq:couplings} lead to a partial decay width
	\begin{equation}
	\Gamma(N \to \overline{udd'}) = \frac{1}{1024\pi^{3}} \left|\frac{y_{d1}y_{u1}}{M_{\Delta 1}^{2}}+\frac{y_{d2}y_{u2}}{M_{\Delta 2}^{2}}\right|^{2}M_{N}^{5}.
	\end{equation}
After baryogenesis there is an excess in $N$ to $\overline{N}$. Without further interactions, and if the $N$ mass is Dirac, the asymmetry in the quarks is exactly erased when the $N$ decay. If the $N$ is Majorana, then both $udd'$ and $\overline{udd'}$ decays are allowed, and the asymmetry can be preserved as long as inverse decays do not become rapid with respect to $H(T = M_{N})$.\footnote{With multiple $N$ generations, CP violation and baryogenesis is also possible via such decays and related scatterings~\cite{Claudson:1983js,Dimopoulos:1987rk,Cheung:2013hza,Baldes:2014rda,Davoudiasl:2015jja,Dev:2015uca,Grojean:2018fus}.} The (inverse) decays are out of equilibrium around $T \approx M_{N}$ if
	\begin{equation}
	\frac{ M_{N} }{ M_{\Delta} } \lesssim 0.3\times \left( \frac{ M_{\Delta} }{ 10^{12} \; \mathrm{GeV} }\right).
	\end{equation}
Then, if the $N$ have no further interactions to remove them from the thermal bath, they will also grow to dominate the energy density of the universe before they decay. However, only a mild dilution of the produced $Y_{B}$ is phenomenologically allowed, given the current baryogenesis mechanism (particularly for AV-type production). Note also that if $M_{N}$ is close to $M_{\Delta}$, there will also lead to suppression of the CP violation in the $\Delta$ decays, due to finite final state mass effects. It is therefore interesting to consider additional interactions for the $N$.

\subsection{Heavy Neutrino for Type-I Seesaw}
An immediate option is to introduce both a Majorana mass $\frac{1}{2}M_{N}\overline{N_{R}}N_{R}^{c}$ and a Yukawa $y_{\nu}\overline{l_{L}}HN_{R}$ term with $y_{\nu}^{2} \sim M_{N}m_{\nu}/v_{\rm EW}^{2}$, and associate $N$ with the type-I seesaw mechanism ($m_{\nu}\sim 0.1$ eV is the light neutrino mass scale and $v_{\rm EW} \sim 100$ GeV is the electroweak scale). One would hope in this way that the massive Majorana Neutrino decays, $N \to H+l$ and $N \to \overline{H}+\overline{l_{L}}$, allow preservation of the asymmetry created during the bubble expansion.\footnote{For $M_{N} \lesssim 10^{9}$ GeV the asymmetry produced in $N$ decays can be naturally suppressed~\cite{Hambye:2001eu,Davidson:2002qv}.} However, one quickly realises the asymmetry created in the $\Delta$ decays is put in jeopardy by $\Delta(L)=2$ scatterings mediated by $N$ around $T \sim M_{N}$. In addition, there are $\Delta(N)=1$ scatterings mediated by SM particles, which are typically in equilibrium already for $T \lesssim 10^{5}M_{N}$~\cite{Giudice:2003jh} and will transfer the negative $B-L$ charge hidden in the $N$ to the active lepton sector, where it can be reprocessed by sphalerons. This means the originally created asymmetry will be erased unless $M_{N} \sim 10^{-5}T_{\rm EW} \sim \mathcal{O}($MeV). But such a low mass scale is difficult to reconcile with BBN as the $N$ are brought into thermal equilibrium in the early universe through their interactions with $\Delta$. Hence the type-I seesaw option for the $N$ is likely ruled out. 

\subsection{Portal to Asymmetric DM}
Finally, one may consider $N$ to decay into a hidden sector, e.g. through a coupling $\sigma\overline{f}N$, where $\sigma$ ($f$) is a dark sector scalar (fermion). If $M_N$ is Majorana, there is no asymmetry preserved in the dark sector, this is a viable option depending on the details of $\sigma$ and $f$, but we do not pursue it further here. If $M_N$ is Dirac, the asymmetry is preserved and the mechanism realises asymmetric DM~\cite{Petraki:2013wwa,Zurek:2013wia}. Furthermore, if there are then sufficiently strong annihilations of the DM particles, the dark matter density is set solely by the asymmetry and mass of the DM particles~\cite{Graesser:2011wi,Iminniyaz:2011yp,Baldes:2017gzw,Baldes:2017gzu}. Barring further entropy injections, the expected masses in this case read $m_{f}+m_{\sigma} \approx 1.8$ GeV. If $\mathrm{Max}[m_f,m_{\sigma}] < m_{n} + m_{\pi}+\mathrm{Min}[m_f,m_{\sigma}]$, where $m_{n} \approx 0.94$ GeV ($m_{\pi} \approx 0.14$ GeV), is the neutron (pion) mass, then the asymmetry erasing DM decay back into the visible sector plus the lightest of the dark sector particles is kinematically forbidden. This translates to $\mathrm{Max}[m_f,m_{\sigma}] \lesssim 1.4$ GeV and $\mathrm{Min}[m_f,m_{\sigma}] \gtrsim 0.4$ GeV. Outside this mass range, the decay can proceed, albeit it is also suppressed, not only by large $M_{\Delta}$, but by the (possibly) high $M_{N}$ mass scale.

\section{Conclusion}
We have begun the study of baryogenesis from heavy particle production during bubble 
expansion. We considered two mechanisms, the mass gain (MG) type production, in which approximately massless particles in the unbroken phase gain a large mass across the bubble wall. We also considered the Azatov and Vanvlasselaer (AV) type production, in which Boltzmann suppressed particles in the unbroken phase are created when light scalar quanta, coupled to the heavy particles, interact with the walls when penetrating the bubble. 

We derived requirements on the phase transition properties, mass spectrum, 
and CP violation for successful generation of the baryon asymmetry. In order to avoid washout, the MG and AV mechanisms work most easily at high scales, with $M_{\Delta} \gtrsim 10^{12}$ GeV. For smaller vacuum energy differences, the MG type production mechanism studied here could in principle also function at lower $M_{\Delta}$ scales. These were two key findings of the current work.

Nevertheless, we wish to emphasise we have not considered resonant enhancement of the CP violation in our current implementation. As is well known from leptogenesis, a resonant enhancement can easily help lower the scale of the decaying particles, while achieving sufficient CP violation and at the same time keeping washout effects suppressed through small Yukawas~\cite{Hambye:2001eu,Davidson:2002qv}. We suspect a qualitatively similar picture may hold in our mechanism which we leave for further exploration. The calculation is not entirely trivial because annihilation effects become increasingly important for small decay rates. Nevertheless, as in known from type-II and III resonant leptogenesis, the required $Y_{B}^{\rm Obs.}$ can still be obtained even in the presence of gauge interactions, because the annihilations quickly become doubly Boltzmann suppressed~\cite{Hambye:2012fh}. Another option would be to implement the ideas in models in which off-shell washout is suppressed by even higher dimensional effective operators than the dimension four appearing in the model considered here.

The gravitational wave amplitude associated to these phase transitions is found to be sizeable, with peak 
frequency $\tilde{f} \approx 330 \, \mathrm{Hz} \, \beta_{H}T_{\rm RH}/10^{10} 
\, \mathrm{GeV}$~\cite{Jinno:2017fby,Konstandin:2017sat,Cutting:2018tjt,Cutting:2020nla,Lewicki:2020jiv}, 
so that the lower mass range of the conventional $c_{\rm vac} \gtrsim 10^{-3}$
parameter space in the AV-type production mechanism can be tested through the IR tail of the gravitational wave signal 
at the Einstein Telescope~\cite{Maggiore:2019uih}. For smaller values of $c_{\rm vac}$ the MG type production can also be tested by the ET.

Some example potentials were also presented that could reproduce the required bulk phase transition parameters for reheating temperatures 
$T_\text{RH} > 10^{13}\,\text{GeV}$, which lead to GW signals outside the ET band. More comprehensive studies of
different microphysical realisations of the effective potential, or resonant enhancements of the CP violation, that would allow one to push $T_\text{RH}$ to lower values, and the associated peak GW frequency into the LISA or ET sensitivity range, are currently being pursued.

\underline{\emph{Acknowledgements.}} ---
IB thanks Thomas Hambye for insightful discussion regarding washout and Kallia Petraki for assistance with colour triplet annihilations. 
IB further thanks St\'ephane Lavignac for a useful question and comment during a workshop discussion which led to some 
refinements of the text. IB is a postdoctoral researcher of the F.R.S.--FNRS with the project `\emph{Exploring new facets of DM}.'
SB, AM, AS and KT are supported by the Strategic Research Program High-Energy Physics and the Research Council of the 
Vrije Universiteit Brussel, 
SB and AM are supported by the ``Excellence of Science - EOS" - be.h project n.30820817 and AS by the FWO-Vlaanderen through the project G006119N. 

\underline{\emph{Note added:}} As this paper was close to being finalised, we became aware that A.~Azatov, M.~Vanvlasselaer, and W.~Yin, were also working on baryogenesis from heavy particle production across relativistic bubble walls~\cite{Azatov:2021irb}. After appearing on the arXiv on the same day, it was realised the two papers are complementary. Reference~\cite{Azatov:2021irb} considers CP violation mostly in the production of the heavy states which can then in principle decay in a CP conserving manner. We have, instead considered the case in which production is necessarily CP conserving, while the decay is CP violating.

\appendix

\section{Leptogenesis style realisation}
\label{sec:LGEN}
In the main text, we have illustrated our mechanism using charged scalars as the decaying particle. Here, we will instead sketch the idea using the heavy Majorana neutrino of the type-I seesaw.
In this case the setup includes a complex field $\Phi$ undergoing the first order phase transition, coupled with Majorana neutrinos as
\be
\mathcal{L} \supset Y_{a,I} \bar L_a H^* N_I + \frac{y_I}{2} \Phi \bar N^c_I N_I + h.c.
\ee
with $I\geq 2$ running on the Majorana neutrino generations and $a$ over the SM lepton generations. 
The neutrinos acquire a mass $M_I \sim y_I \langle \Phi \rangle$ in the true vacuum, while they are massless in the false vacuum and in thermal equilibrium
with the SM.
For sufficiently boosted bubbles $\gamma T_n > M_I $, the heavy neutrinos are generated 
through the bubble with yield 
$Y_{N_I}(T_n) \sim Y^{\text{eq}}_{N_I} \left(T_n/T_{RH} \right)^3$. 
The asymmetry is generated through the CP violating decays of the heavy neutrinos $N_I \to H^* \bar L_{a}$.
In \cite{Shuve:2017jgj} a similar scenario was investigated, but the possibility that the heavy neutrinos are created thanks to boosted bubbles not explored.

This leptogenesis style scenario presents  
several differences with the MG case discussed in the main body of the paper: 
i) The phase transition can be induced by gauging the $U(1)_{B-L}$ symmetry under which $\Phi$ is charged, in a classical scale invariant model~\cite{Marzo:2018nov}, 
leading to supercooling and boosted bubbles (so no other scalar degrees of freedom are needed);
ii) The decay products of the heavy particle $N_I$, that is $H$ and $L_a$, possess EW interactions and hence reach fast thermal equilibrium with the SM plasma after wall crossing,
thus alleviating possible washout processes like the ones discussed in Sec.\,\ref{sec:avoidingwashoutA};
iii) The washout processes after reheating benefit of a suppression $\sim \left( T_{RH}/M_{I} \right)$
analogously to what discussed in Sec.\,\ref{sec:avoidingwashoutB} for $\Delta$, so we do not expect qualitative changes there.

These observations make this setup another promising scenario to realise bubble baryogenesis, in addition to the models illustrated in the main text.
We leave a detailed investigation of this realisation for future studies.

\section{Estimate of the boosted on-shell washout.}
\label{app:OSWO}
Consider a boosted particle produced by the $\Delta$ decay inside the bubble. In the plasma frame. we take its four momentum to be
	\begin{equation}
	p_{\ast}^{\mu} = E_{\ast}(1,0,0,1) \equiv p_{\ast}(1,0,0,1).
	\end{equation}
To match our notation in the main text we will later write $E_{\ast} = M_{\Delta}^{2}/(kT)$, where $k \approx 2^{J}$ corresponds to the $J$'th scattering step (when notationaly convenient, we switch subscript $\ast$ to superscript $\ast$ below, to keep track of which is the boosted particle). Consider now, for concreteness, the initial particle to be a boosted right chiral up quark. We can write the initial boosted phase space distribution as
	\begin{equation}
	\label{eq:boostdist}	
	f^{\ast}_{u} = n_{\ast}^{u}(2\pi)^{3} \delta(p_x)\delta(p_y)\delta(p_z-p_{\ast}),
	\end{equation}
where for simplicity we consider only a monochromatic distribution. In addition, there would, of course, also be a distribution of $\overline{u_{R}}$, with $n_{\ast} -n_{\ast}^{\bar{u}} \simeq n_{B}$. Now we consider the interaction of the boosted $u_{R}$'s with unboosted $N_{R}$'s in the plasma frame. The latter have a thermal distribution which we approximate with the massless Maxwell-Boltzmann one
	\begin{equation}
	\label{eq:MBeq}
	f_{N}(p) = e^{-E(p)/T} = e^{-p/T}. 
	\end{equation} 
Now we wish to approximate the rate of inverse decays onto on-shell states, i.e.~$uN \to \Delta$. This rate is given by
	\begin{align}
	\frac{dn_{\rm Inv}}{dt} & = \int \frac{d^{3}p_{N}}{(2\pi)^{3}2E_{N}}\int \frac{d^{3}p_{\Delta}}{(2\pi)^{3}2E_{\Delta}} \int  \frac{d^{3}p_{u}^{\ast}}{(2\pi)^{3}2E_{u}^{\ast}} \nonumber \\
				& \quad \times (2\pi)^{4}\delta^{4}(p_u^{\ast} + p_N - p_{\Delta}) f^{\ast}_{u} f_{N} |\mathcal{M}|^{2},
	\end{align}
where $|\mathcal{M}|^{2} = 2|y_{u1}|^{2}(p_u^{\ast})_{\mu}(p_N)^{\mu}$ is the matrix element squared (here in the CP even limit for simplicity). Using the delta functions in Eq.~\eqref{eq:boostdist} to do the $d^{3}p_{u}^{\ast}$ integrals, and substituting in for Eq.~\eqref{eq:MBeq},  one finds
	\begin{align}
\frac{dn_{\rm Inv}}{dt} & = \frac{n_{\ast}^u}{2E_{\ast}} \int \frac{d^{3}p_{N}}{(2\pi)^{3}2E_{N}} \int \frac{d^{3}p_{\Delta}}{(2\pi)^{3}2E_{\Delta}} \nonumber \\
			&  \times (2\pi)^{4}\delta^{4}(p_u^{\ast} + p_N - p_{\Delta}) e^{-E_N/T} |\mathcal{M}|^{2}.
	\end{align}
Next we wish to do the $d^{3}p_{\Delta}$ integral over the three momentum delta functions. This will set $\vec{p}_{\Delta} = \vec{p}_u^{\ast}+\vec{p}_{N}$ and given that the out-going $\Delta$ is on-shell, 
	\begin{equation}
	E_{\Delta}[E_N,c_\theta] = \sqrt{ M_{\Delta}^{2} + E_{\ast}^{2} + E_{N}^{2} + 2E_{\ast}E_N c_\theta },
	\end{equation}
where $c_\theta \equiv \mathrm{cos}(\theta)$ is the cosine of the angle between $ \vec{p}_u^{\ast}$ and $\vec{p}_{N}$. Hence we have
	\begin{align}
	\frac{dn_{\rm Inv}}{dt} & = \frac{n_{\ast}^u}{16 \pi E_{\ast}} \int dc_{\theta} \int \frac{ dE_N E_N e^{-E_N/T}|\mathcal{M}|^{2} }{ E_{\Delta}[E_N,c_\theta] } \nonumber \\ & \qquad  \times  \delta(E_{\ast}+E_N - E_{\Delta}[E_N,c_\theta]).
	\end{align}
The root of the function inside the remaining $\delta$ is at $E_{N}^{0} = M_{\Delta}^{2}/2E_{\ast}(1-c_{\theta})$. Using the properties of the delta function we can perform the energy integral and find
	\begin{equation}
	\frac{dn_{\rm Inv}}{dt} = \frac{n_{\ast}^u}{16 \pi E_{\ast}}  \int dc_{\theta} \frac{E_{N}^{0} e^{-E_{N}^{0}/T}|\mathcal{M}|^{2}}{|E_{\Delta}[E_{N}^{0},c_\theta]-E_{N}^{0}-E_{\ast}c_{\theta}|}.
	\end{equation}
Evaluating the angular integral using Mathematica and substituting in $|\mathcal{M}|^{2} = |y_{u1}|^{2}M_{\Delta}^2$,  we find
	\begin{equation}
	\frac{dn_{\rm Inv}}{dt}  = \Gamma_{1u}\frac{ M_{\Delta} T_n }{ E_{\ast}^{2} } \times \mathrm{Exp} \left[ -\frac{ M_{\Delta}^{2} }{ 4E_{\ast}T_n } \right]. 
	\end{equation}
For the washout rate, this expression should be multiplied by the probability of the produced $\Delta$ decaying in such a way that the overall process violats $B-L$. One should also subtract the rate of the initial antiparticle state (or, essentially equivalently, just consider the rate per initial particle). We therefore arrive at the required  effective $B-L$ violating inverse decay rate
	\begin{align}
 	\Gamma_{\rm Inv} & = \frac{1}{n_{\ast}^u}\frac{dn_{\rm Inv}}{dt}\times \mathrm{Br}(\Delta \to \overline{d'}\overline{d}) \\
	& = \frac{ \Gamma_{1u}\Gamma_{1d} }{ \Gamma_{\Delta} } \frac{ M_{\Delta} T_n }{ E_{\ast}^{2} } \times \mathrm{Exp} \left[ -\frac{ M_{\Delta}^{2} }{ 4E_{\ast}T_n } \right],
	\end{align}
which is the result stated in Eq.~\eqref{eq:appinv}.

\section{Boltzmann equations for the thermal washout}
\label{app:Bmann}
In the main text, we have estimated the washout at $T_{\rm RH}$ using simple dimensional estimates. Here we provide detailed Boltzmann equations for the washout which allow us to study the effect with greater accuracy. Consider the $(B-L)$ violating interactions introduced in Eq.~\eqref{eq:couplings}. We consider the following decays and scatterings
	\begin{align}
	\Delta & \leftrightarrow \overline{d}+\overline{d'} \\
	\Delta^{\ast} & \leftrightarrow \overline{u} + \overline{N} \\
	d+d' & \leftrightarrow \overline{u}+\overline{N} \\
	d+u  & \leftrightarrow \overline{d'}+\overline{N}
	\end{align}
plus processes related to the above by CP and T reversal. The decays have been defined and evaluations given in the main text. For the scatterings, we introduce the following reaction rate densities
	\begin{align}
	\gamma_{1} & \equiv \frac{n_{d}^{\rm eq}n_{d'}^{\rm eq}}{g_{d}g_{d'}}\langle v_{\rm rel} \sigma_{dd' \to \overline{u}\overline{N}} \rangle \\
	\gamma_{2} & \equiv \frac{n_{d}^{\rm eq}n_{u}^{\rm eq}}{g_{d}g_{u}}\langle v_{\rm rel} \sigma_{du \to \overline{d'}\overline{N}} \rangle,
	\end{align}
where we have factored out the usual degrees-of-freedom in order to more clearly be able to sum over initial and final states in our equations below. The reaction rate densities can be evaluated through
	\begin{equation}
	\gamma = \frac{T}{8\pi^{4}} \int_{(m_a+m_b)^{2}}^{\infty}ds E_{a}E_{b}p_{i} v_{\rm rel}\sigma K_{1}\left(\frac{\sqrt{s}}{T} \right),
	\label{eq:ratedensity}
	\end{equation}
where $m_{a}$ and $m_{b}$ ($E_{a}$ and $E_{b}$) are the initial particle masses (energies), $p_{i}$ is the initial centre-of-mass momentum.

The Boltzmann equation for the asymmetry is then given by
	\begin{align}
	\frac{dn_{B-L}}{dt} + & 3Hn_{B-L}  = \nonumber \\
	\sum_{\rm initial}  \sum_{\rm final} & \bigg\{ \, \gamma_{1}\left( \frac{ \overline{ n_{d} n_{d'}} }{ n_{d}^{\rm eq}n_{d'}^{\rm eq} } - \frac{ n_{d} n_{d'} }{ n_{d}^{\rm eq}n_{d'}^{\rm eq} } + \frac{ \overline{n_{u}n_{N}} }{ n_{u}^{\rm eq}n_{N}^{\rm eq} } - \frac{ n_{u}n_{N} }{ n_{u}^{\rm eq}n_{N}^{\rm eq} } \right) \nonumber \\
	& \; + \gamma_{2} \left( \frac{ \overline{n_{d}n_{u}} }{ n_{d}^{\rm eq}n_{u}^{\rm eq} } - \frac{ n_{d}n_{u} }{ n_{d}^{\rm eq}n_{u}^{\rm eq} } + \frac{ \overline{n_{d'}n_{N}} }{ n_{d'}^{\rm eq}n_{N}^{\rm eq} } - \frac{ n_{d'}n_{N} }{ n_{d'}^{\rm eq}n_{N}^{\rm eq} }  \right) \nonumber \\
	& \; + \frac{2}{3}\Gamma_{1}n_{\Delta}^{\rm eq}\left( \frac{ \overline{n_{\Delta}} }{ n_{\Delta}^{\rm eq} }  -  \frac{ n_{\Delta} }{n_{\Delta}^{\rm eq}} + \frac{\overline{ n_{d} n_{d'}} }{ n_{d}^{\rm eq}n_{d'}^{\rm eq} } - \frac{ n_{d} n_{d'} }{ n_{d}^{\rm eq}n_{d'}^{\rm eq} } \right) \nonumber \\ 
	& \; + \frac{1}{3}\Gamma_{2}n_{\Delta}^{\rm eq}\left(  \frac{ n_{\Delta} }{n_{\Delta}^{\rm eq}} - \frac{ \overline{n_{\Delta}} }{ n_{\Delta}^{\rm eq} } + \frac{ \overline{n_{u}n_{N}} }{ n_{u}^{\rm eq}n_{N}^{\rm eq} } - \frac{ n_{u}n_{N} }{ n_{u}^{\rm eq}n_{N}^{\rm eq} } \right) \nonumber \\
	& + \epsilon_{\Delta}\Gamma_{\Delta}\left( n_{\Delta} + \overline{n_{\Delta}} - 2n_{\Delta}^{\rm eq} \right) \bigg\}, \label{eq:Bmann}
	\end{align}
where we have also included the CP violating source term for decays. For this last term, we have used the unitarity relation between the CP asymmetry in the decay and the CP asymmetry in the $2 \leftrightarrow 2$ scatterings with the on-shell intermediate state subtracted, which leads to no asymmetry being generated in equilibrium. (In principle there can also be a additional CP violating source terms for $2 \leftrightarrow 2$ scatterings, coming away from the $\sqrt{s} \simeq M_{\Delta}$ region, depending on the number of quark and $N$ generations taking part. This gives an additional non-zero source term when, for example, $N$ is out-of-equilibrium. We do not write it here.)

We now wish to study washout when a $Y_{B-L}$ is present. To do this, we relate the ratios of number densities over their equilibrium values to the chemical potentials of the individual species when $Y_{B-L}\neq 0$. Assuming quarks and leptons reach chemical equilibrium for simplicity, we have
	\begin{equation}
	\frac{\mu_{d_R}}{T} = \frac{38}{79} \frac{s_e}{T^{3}} Y_{B-L}, \qquad \frac{\mu_{u_R}}{T} = -\frac{10}{79} \frac{s_e}{T^{3}} Y_{B-L}.
	\end{equation} 
(The SM Yukawas are not sufficient for the light leptons to be in chemical equilibrium at high $T$, but additional BSM interactions may ensure this. Above $T \sim 10^{12}$ GeV, electroweak sphaleron mediated processes are also out of equilibrium.) For the $N_{R}$ we assume a single generation, which does not share its asymmetry with a hidden sector, so that
	\begin{equation}
	\frac{\mu_{N_R}}{T} = 3 \frac{s_e}{T^{3}} Y_{B-L}.
	\end{equation}
Realistically, $N_{R}$ would share some of its asymmetry, which would reduce the above chemical potential and also reduce the washout. The above Boltzmann equation can be recast in the form
	\begin{equation}
	\frac{dY_{B-L}}{dT} = - \sqrt{ \frac{\pi}{45} } \frac{ M_{\rm Pl}g_{\ast}^{1/2} }{ s_e^{2} } \times C, 
	\end{equation}
where $C$ is the collision term, i.e.~the right-hand-side of Eq.~\eqref{eq:Bmann}. For simplicity, we go on to assume that only one quark flavour combination $dd'u$ is contributing in the Boltzmann equation. Then focusing solely on the scattering terms and linearizing with respect to the small chemical potentials, we have
	\begin{equation}
	\frac{dY_{B-L}}{dT} = - \sqrt{ \frac{45}{4\pi^{3}} } \frac{ M_{\rm Pl} Y_{B-L}}{g_{\ast}^{1/2}T^{6}} \left( \frac{606}{79} \right) (\gamma_{1}+2\gamma_{2}), 
	\label{eq:WOsingleflav}
	\end{equation}
where the factor of two in front of $\gamma_{2}$ comes from a sum over flavours. 

Let us now evaluate the required cross sections and reaction rate densities with the simplifying assumption of massless quarks (i.e.~we ignore their thermal masses). In order to study the scatterings at $T_{\rm RH} \ll M_{\Delta}$, We integrate out the heavy $\Delta$ and write an effective coupling,
	\begin{equation}
	\mathcal{L} \supset \frac{y^{2}}{M_{\Delta}^{2}}\epsilon_{ijk}(\overline{d_{Ri}^{c}}d_{Rj}')(\overline{u_{Rk}^{c}}N_{R}) + \mathrm{H.c.}
	\end{equation}
where $i,j,k$ are colour indices. Then summing over initial and final state spins and colour possibilities (the latter introduces an overall multiplicative factor of six for both cross sections --- and we have already summed over flavours above), we find
	\begin{align}
	E_{d}E_{d'}p_{i} v_{\rm rel}\sigma_{dd' \to \overline{u}\overline{N}} = \frac{3}{32\pi} \frac{y^{4}}{M_{\Delta}^4}s^{5/2}, \\
	E_{d}E_{u}p_{i} v_{\rm rel}\sigma_{du \to \overline{d'}\overline{N}} = \frac{1}{32\pi} \frac{y^{4}}{M_{\Delta}^4}s^{5/2}.
	\end{align}
We plug these into Eq.~\eqref{eq:ratedensity} and find
	\begin{equation}
	\gamma_{1} = 3\gamma_{2} = \frac{9 \, y^{4} T^{8}}{\pi^{5}M_{\Delta}^4}.
	\end{equation}
As we are working in an EFT, this is valid for $T \ll M_{\Delta}$, which ensures the Bessel function in Eq.~\eqref{eq:ratedensity} effectively cuts off the integral before the approximation for the cross section breaks down. In practice, we anyway need $T_{\rm RH} \lesssim M_{\Delta}/10$ to realistically avoid washout, so the EFT approximation is good. We can substitute this into Eq.~\eqref{eq:WOsingleflav} and re-arrange to find
	\begin{equation}
	\frac{d \log(Y_{B-L})}{d \log(T) } \approx 0.23 \frac{y^{4}M_{\rm Pl} T^{3}}{M_{\Delta}^{4}g_{\ast}^{1/2}}.
	\label{eq:BMannWOprecise}
	\end{equation}
Demanding the right-hand-side is less than unity to avoid washout is, up to an $\mathcal{O}(1)$ factor, equivalent to dividing Eq.~\eqref{eq:WOguess} by $H$ at $T_{\rm RH}$ and demanding the ratio is below one. We have of course, also evaluated Eq.~(\ref{eq:BMannWOprecise}) numerically as a check, and displayed the results in Fig.~\ref{fig:WO}.

\bibliography{bbb}
\end{document}